\documentclass[final,3p,11pt]{elsarticle} 

\usepackage{amssymb}
\usepackage{amsmath}
\usepackage{siunitx}
\usepackage{algorithm}
\usepackage{algorithmic}
\usepackage{float} 
\usepackage{tikz}
\usepackage{graphicx}
\usepackage{subcaption}
\journal{Placeholder Jrnl}
\usepackage{hyperref}
\usepackage{natbib}

\begin{document}

\begin{frontmatter}

\title{Efficient spectral Galerkin framework for nonlinear transient heat transfer in finite domains}

\author{Théo Andrieux}
\author{Andreas Ntinos}
\author{Manas V. Upadhyay\corref{cor1}}
\affiliation{organization={Laboratoire de Mécanique des Solides (LMS), École Polytechnique, CNRS UMR 7649, Institut Polytechnique de Paris},
            addressline={Route de Saclay},
            city={Palaiseau},
            postcode={91128},
            country={France}}
\ead{manas.upadhyay@polytechnique.edu}
\cortext[cor1]{Corresponding author}

\begin{abstract}
Accurately modelling the temporal evolution of heterogeneous temperature fields requires resolving strong nonlinearities in the heat equation arising from temperature-dependent thermophysical properties, latent heats of transformation and any local heat sources/sinks. 
In this work, we present a spectral Galerkin (SG) framework to solve the fully nonlinear transient heat equation in finite domains to attain the accuracy of high-fidelity finite element (FE) simulations at considerably lower computational cost.
The heat equation is reformulated into a linear reference problem with constant thermophysical properties and residual forcing terms. 
Solving the reference problem provides a complete three-dimensional orthonormal trigonometric basis, whose Galerkin projection reduces the heat equation to a set of modal ordinary differential equations (ODEs) in time; the reference operator is diagonal in the modal basis, and results in independent modal updates for a fixed nonlinear forcing. 
These ODEs can be integrated using exponential time differencing and iteratively corrected for nonlinearities.
The SG method eliminates the global solve required by FE methods, and its use of structured grids allows efficient GPU parallelization.
Applied to rapid laser–metal interactions, the SG solver reproduces high-fidelity FE temperature fields with less than 1\% relative error while achieving 227-fold faster GPU runtimes.
Applied to a part-scale laser scanning study [Ramani et al., Additive Manufacturing 52 (2022) 102643], the method shows that accounting for evaporation and latent heat more than 
halves their proposed processing metric.
The source code of the SG heat solver and some worked examples are available at \url{https://github.com/manasvupadhyay/spectral_galerkin_heat} under the Apache 2.0 license.
\end{abstract}


\end{frontmatter}

\section{Introduction}

Nonlinear transient heat transfer occurs across a wide range of thermal processes involving phase transformations, strongly temperature-dependent material properties, localized heat sources or sinks, and nonlinear boundary interactions.
Accurate resolution of such problems generally relies on high-fidelity approaches such as finite element (FE), finite volume, and finite difference formulations, which can accommodate complex nonlinear behaviour and boundary conditions but require the repeated solution of large matrix problems.
Their computational cost becomes particularly expensive when strong spatial gradients, short characteristic time scales, or repeated simulations require fine spatial and temporal discretizations. 

Analytical and semi-analytical approaches provide an attractive alternative because of their significantly lower computational cost, but these solutions are derived for linear operators with constant coefficients and relatively simple boundary conditions \cite{carslawConductionHeatSolids1947,ozisikHeatConduction1993}.
Furthermore, widely used solutions for moving heat sources often rely on infinite or semi-infinite domains, as demonstrated by the Rosenthal and Eagar--Tsai solutions \cite{rosenthal1946theory,eagar1983temperature}. 
A considerable gap therefore remains between computationally intensive high-fidelity numerical approaches and rapid  analytical solutions, when nonlinear thermal evolution must be resolved in finite domains.
In this work, we use several independently existing and established mathematical and numerical concepts as building blocks to bridge this gap. 
The starting point is the decomposition of the nonlinear transient heat transfer problem, including the governing equation and boundary conditions, into a linear isotropic reference problem and a residual forcing term. 
The linear isotropic volumetric and surface reference operators are endowed with constant thermophysical properties.
Variations with respect to the linear isotropic reference operators, including temperature-dependent density and heat capacity, spatially and thermally varying anisotropic conductivity, latent heats of transformation, volumetric heat sources or sinks, and nonlinear surface fluxes are taken as residual forcing.
This decomposition is motivated by fast Fourier transform based numerical homogenization techniques \cite{moulinecFastNumericalMethod1994a,moulinecNumericalMethodComputing1998,michelComputationalSchemeLinear2001}, where the heterogeneous constitutive problem is reformulated about a homogeneous reference medium and material fluctuations enter through a polarization field.
The reference problem involving the linear isotropic volumetric and surface operators, together with homogeneous boundary conditions, defines the classical eigenvalue problem whose eigenfunctions can be obtained analytically for separable geometries such as a cuboid \cite{carslawConductionHeatSolids1947,habermanAppliedPartialDifferential2013}.
The reference Laplacian operator on a bounded domain with homogeneous boundary conditions (Neumann, Dirichlet or Robin-type) is self-adjoint with a compact resolvent, and therefore it provides a complete orthonormal eigenbasis for the admissible function space \cite{evansPartialDifferentialEquations2010}. 
Expanding the temperature field on this basis and applying a Galerkin projection transforms the spatially discretized problem into a system of modal ordinary differential equations (ODEs) in time \cite{canutoSpectralMethodsEvolution2007,hesthavenSpectralMethodsTimeDependent2007}.
These ODEs can then be integrated analytically using exponential time-differencing methods \cite{coxExponentialTimeDifferencing2002,kassamFourthOrderTimeStepping2005}, while nonlinear volumetric and surface terms can be updated iteratively e.g., through a fixed-point procedure.

These concepts are combined together for the first time in this work to form a unified framework, called the Spectral Galerkin (SG) framework, for nonlinear, transient, and generally anisotropic heat transfer problems in finite domains.
The SG framework retains all the nonlinearities while providing the computational advantages of a diagonal spectral reference operator; this eliminates the need for repeated global algebraic solves that are characteristic of implicit FE formulations.
The resulting structured-grid formulation is also well suited for massively parallel GPU implementation.

The SG framework is applied to study temperature evolution during rapid laser--metal (316L austenitic stainless steel) interactions with melting and evaporation, which are considered as a representative set of applications as they combine many of the nonlinearities that make conventional FE simulations computationally very demanding.
A translating localized heat source generates strong three-dimensional temperature gradients over short time scales. 
Melting introduces latent heat of fusion and important variations in thermophysical properties in the liquid and solid, and evaporation can produce a highly temperature-dependent surface heat loss \cite{kingLaserPowderBed2015,debroyAdditiveManufacturingMetallic2018,matthewsDenudationMetalPowder2016}.
The predictions of the SG formulation are assessed against a high-fidelity nonlinear implicit FE solver to study the trade-off between computational gains and accuracy in the temperature field. 
The method is then applied to the part-scale laser scanning strategy problem considered by Ramani et al.~\cite{ramaniSmartScanIntelligentScanning2022}. 
In their study, Ramani et al. employed a finite difference solver reduced with radial basis functions and used linear coefficients and a Gaussian heat source to construct a temperature-based processing metric. 
The proposed SG framework additionally accounts for latent heat, evaporation, and temperature-dependent thermophysical properties, while keeping the computational times practical, and it is used to quantify the differences with the results in \cite{ramaniSmartScanIntelligentScanning2022}. 
The resulting differences in the predicted temperature field and processing metric are used to demonstrate the significance of resolving nonlinear effects in rapid thermal-process optimization.

The remainder of the article is organized as follows. 
Section~2 formulates the general nonlinear anisotropic heat-transfer problem and its volumetric and surface contributions on a cuboid domain with Neumann boundary conditions. 
Section~3 develops the semi-analytical SG framework of this problem, including construction of the linear isotropic reference operators, derivation of the homogeneous eigensystem, and Galerkin projection onto the resulting spectral basis, exponential time integration, fixed-point treatment of the nonlinear forcing terms, and the algorithm.
Note that the same steps can be followed to develop the framework for cylindrical and spherical domains, and for different boundary condition combinations involving Dirichlet and Robin-type conditions; however, these developments are not pursued in this work.
Section~4 describes the simulation setup for the rapid laser--metal interaction problem and the SG and FE numerical implementations.
Section~5 assesses the accuracy and computational performance of the SG framework against analytical and FE reference solutions before applying it to rapid laser scanning and to the scan-strategy problem of Ramani et al.~\cite{ramaniSmartScanIntelligentScanning2022}. 
Section~6 discusses the origin of the computational advantage of the SG approach over the FE one, assesses its limitations and presents potential improvements to be undertaken. 
The main findings and perspectives for extending the framework are summarized in the conclusion section.
The appendix presents the FE algorithm. 
Finally, the source code for the SG implementation with the worked examples is available at this \href{https://github.com/manasvupadhyay/spectral_galerkin_heat}{link} under the Apache 2.0 licence.

\section{Nonlinear heat transfer in finite domains}\label{sec:problem}

A three-dimensional (3D) cuboid-shaped domain $\Omega = [0, L_x] \times [0, L_y] \times [0, L_z]$ with surface $\partial \Omega$ is considered in a Cartesian basis; the same developments carry over to cylindrical and spherical domains. 

The governing equation of the heat transfer problem is taken as:
\begin{equation}\label{eq:heat_equation}
     \rho\left[ c + \sum_j L_j \frac{\partial f_j}{\partial T} \right] \dot{T} = \nabla \cdot (\boldsymbol{K} \cdot \nabla T) + S \quad \text{in } \Omega \times \left(0,t_F\right]
\end{equation}
where $T \equiv T(\boldsymbol{x},t)$ is the temperature field and the overhead dot represents the partial time derivative $\partial/\partial t$. 
$\rho\equiv \rho(T)$ is the temperature-dependent mass density. $c \equiv c(T) = \sum_{i} c_i(T) \chi_i$ captures the specific heat capacity $c_i(T)$ of all the $i$ phases with $\chi_i$ mass fraction such that $\sum_i \chi_i = 1$. $L_j$ represents the specific latent heat of the $j$-th phase transition and 
$f_j \equiv f_j(T)$, for $T \in [T_{s,j}, T_{e,j}]$, is the associated transformed fraction with $T_{s,j}$ and $T_{e,j}$ being the start and end temperatures, respectively. 
$\nabla$ is the gradient operator.
$\boldsymbol{K}\equiv \boldsymbol{K}(T)$ is the generally anisotropic temperature-dependent second-order thermal conductivity tensor. 
$S \equiv S(\boldsymbol{x},t,T)$ is a heat source term that collects contributions from other dissipative mechanisms e.g., chemical reactions, thermomechanical couplings, etc.
$t_F$ is the final simulation time.
Note that the Fourier law of heat conduction $\boldsymbol{q} = -\boldsymbol{K} \cdot \nabla T$, where $\boldsymbol{q}$ is the heat flux vector, is implicitly used in \eqref{eq:heat_equation}. 

Neumann boundary conditions are prescribed on the entire surface $\partial \Omega$ and they are expressed as the sum of individual heat flux contributions:
\begin{equation}\label{eq:boundary_flux}
-\left( \boldsymbol{K} \cdot \nabla T \right) \cdot \boldsymbol{n} = \sum_{s=1}^{N_s} q_s(\boldsymbol{x},t, T) \quad \text{on } \partial \Omega \times (0,t_F]
\end{equation}
where $\boldsymbol{n}$ is the normal to a surface.
$q_s$ defines the $s$-th surface heat flux due to various physical phenomena such as moving heat source and heat losses to the surroundings (convective, evaporative and radiative); each phenomenon is represented by a distinct flux, possibly nonlinear in temperature.
The equations, derivations, and numerical examples are presented only with Neumann boundary conditions; the same procedures, however, extend to other conditions, such as Dirichlet or Robin-type (mixed Neumann and Dirichlet) conditions.

The general initial condition is a spatially varying temperature field:
\begin{equation}\label{eq:init_cond}
    T(\boldsymbol{x},t=0) = T_{\text{init}}(\boldsymbol{x}) \quad \text{in } \Omega
\end{equation}

\section{Semi-analytical spectral Galerkin (SG) framework}
\label{sec:sg_method}

To construct a semi-analytical solution, the nonlinear heat problem introduced in Section~\ref{sec:problem} is reformulated by separating the governing equations into a linear isotropic reference problem and a collection of nonlinear volumetric and surface forcing terms. 
The reference problem is intentionally chosen such that its governing operator is linear, isotropic, and subject to homogeneous Neumann boundary conditions. 
This enables the construction of a homogeneous eigenproblem whose eigensystem depends solely on the computational geometry and the homogeneous boundary conditions, and is therefore independent of the nonlinear forcing terms. 
Since the reference eigenproblem is linear and separable on a finite cuboidal domain, its eigensystem can be obtained analytically using the classical method of separation of variables. 
The eigensystem constitutes a complete orthonormal basis for the subsequent spectral expansion and Galerkin projection, which reduces the governing partial differential equation to a system of modal ODEs in time.
These equations are solved with an iterative scheme that lags the nonlinear volumetric and surface forcing terms together, evaluating them at the previous iterate and converging to a self-consistent solution within each time step with a fixed point iterative loop.

\subsection{Construction of a linear isotropic reference operator}
\label{sec:decomposition}

Motivated by linear reference problems in mechanics \cite{moulinecFastNumericalMethod1994a, moulinecNumericalMethodComputing1998, eyreFastNumericalScheme1999} 
the nonlinear heat transfer problem described in Section~\ref{sec:problem} is first rearranged to obtain a linear isotropic reference operator.
This operator is selected to retain the dominant constant-coefficient storage and diffusion terms, while fluctuations arising from temperature-dependent material properties, anisotropy, latent heat and nonlinear boundary conditions are treated as forcing terms.

Each temperature-dependent thermophysical property can be split into a constant reference value and a fluctuation about it,
\begin{equation}\label{eq:property_split}
    \begin{aligned}
        \rho(T) &= \overline{\rho} + \widetilde{\rho}(T) \\
        c(T) &= \overline{c} + \widetilde{c}(T) \\
        \boldsymbol{K}(T) &= \overline{k} \, \mathbb{I} + \widetilde{\boldsymbol{K}}(T)
    \end{aligned}
\end{equation}
where $\overline{\rho}$, $\overline{c}$ and $\overline{k}$ are constant reference values, $\widetilde{\rho}$, $\widetilde{c}$ and $\widetilde{\boldsymbol{K}}$ the fluctuations about them, and $\mathbb{I}$ the second-order identity tensor. 
The magnitude of the reference constants strongly affects the magnitude of the forcing term, and a convergence analysis may be performed during a numerical simulation to determine a suitable choice. 

Linear isotropic reference bulk $\mathcal{L}$ and surface $\mathcal{L}_s$ operators are defined as:
\begin{equation}
    \begin{aligned}
        \mathcal{L} :=& \ \overline{\rho} \, \overline{c} \, \frac{\partial}{\partial t} - \overline{k} \nabla^2 \\
        \mathcal{L}_s :=& -\overline{k} \, \boldsymbol{n} \cdot \nabla
    \end{aligned}
\end{equation}
where $\nabla^2$ is the Laplacian.

Then, the nonlinear heat equation and the Neumann boundary conditions can be exactly rewritten using these operators as
\begin{eqnarray}
    \mathcal{L} T &=& R(T) \quad \text{in } \Omega \times (0,t_F)\label{eq:Lheatproblem} \\
    \mathcal{L}_sT &=& R_s(T) \quad \text{on } \partial \Omega \times  (0,t_F)\label{eq:Lheatproblem_surf}
\end{eqnarray}
where
\begin{eqnarray}
        R(T) &=& \nabla \cdot \left( \widetilde{\boldsymbol{K}} \cdot \nabla T \right) - \rho \sum_j L_j \frac{\partial f_j}{\partial T} \dot{T} - \left( \overline{\rho} \, \widetilde{c} + \widetilde{\rho} \, \overline{c} + \widetilde{\rho} \, \widetilde{c} \right) \dot{T} + S \quad \text{and} \\
        R_s(T) &=& \sum_{s=1}^{N_s} q_s + \left( \widetilde{\boldsymbol{K}} \cdot \nabla T \right) \cdot \boldsymbol{n}\label{eq:Rs}  
\end{eqnarray}
are treated as bulk and surface forcing terms, respectively, which contain all those terms that prevent obtaining an analytical solution to the problem.


\subsection{Spatially homogeneous reference problem and its eigenfunctions}

Reformulating the nonlinear heat problem as shown in equation~\eqref{eq:Lheatproblem} enables the construction of a homogeneous reference eigenproblem associated with the linear isotropic reference operator. 
The resulting eigenfunctions depend solely on the geometry and homogeneous boundary conditions, and are therefore independent of the nonlinear volumetric and surface forcing terms.
Owing to the linearity of the reference operator and the homogeneous Neumann boundary conditions, the resulting eigenfunctions can be obtained analytically by the method of separation of variables. 

The homogeneous reference eigenproblem is posed as:
\begin{equation}\label{eq:homogeneouseigen}
    \begin{aligned}
        \mathcal{L}T =& \ 0 \quad &\text{in } \Omega \times (0,t_F] \\
    \mathcal{L}_sT =& \ 0 \quad &\text{on } \partial \Omega \times (0,t_F]
    \end{aligned}
\end{equation}

Following the method of separation of variables, let $T(\mathbf{x}, t) = \Phi(\mathbf{x}) \Theta(t)$ and substitute into \eqref{eq:homogeneouseigen} to obtain:
\begin{equation}
    \frac{\overline{\rho} \,  \overline{c}}{\overline{k}} \frac{\dot{\Theta}}{\Theta} = \frac{\nabla^2 \Phi}{\Phi} = -\lambda
\end{equation}
where the separation of variables implies that $\lambda$ is a constant in both space and time.

The spatial part yields an eigenvalue problem for the Laplacian operator with purely Neumann boundary conditions.

\begin{equation}\label{eq:laplaceproblem}
\begin{aligned}
    -\nabla^2 \Phi =& \ \lambda \Phi  \quad &\text{in } \Omega  \\
    \nabla\Phi \cdot \boldsymbol{n} =& \ 0 \quad &\text{on } \partial \Omega
\end{aligned}
\end{equation}

Let $\Phi(x,y,z) = X(x)Y(y)Z(z)$. Substituting it into the spatial eigenvalue problem and dividing yields:

\begin{equation}
-\frac{X''}{X} - \frac{Y''}{Y} - \frac{Z''}{Z}= \lambda
\end{equation}
where each prime represents a total derivative.

Since $X$, $Y$ and $Z$ are independent of each other and $\lambda$ is a constant, each term on the left hand side of this equation must equal to a constant, say $\lambda_x^2$, $\lambda_y^2$ and $\lambda_z^2$, respectively, such that $ \lambda = \lambda_x^2 + \lambda_y^2 + \lambda_z^2$.

This 3D problem can then be decoupled into three independent one-dimensional Sturm-Liouville problems in $x$, $y$ and $z$ as
\begin{equation}
\begin{aligned}
    - X'' =& \lambda_x^2 X, \quad 
    &X'(0) =& \, X'(L_x) = 0 \\
    - Y'' =& \lambda_y^2 Y, \quad 
    &Y'(0) =& \, Y'(L_y) = 0 \\
    - Z'' =& \lambda_z^2 Z, \quad 
    &Z'(0) =& \, Z'(L_z) = 0
\end{aligned}
\end{equation}

The admissible solutions to these independent problems yield the following one-dimensional eigenfunctions:
\begin{equation}
\begin{aligned}
    \Phi_m =& C_m \cos(\lambda_x^m x) \quad &\text{with } \lambda_x^m =& \ m \pi/L_x ; \quad &m = 0, 1, 2, ... \\
    \Phi_n =& C_n \cos(\lambda_y^n y) \quad &\text{with } \lambda_y^n =& \ n \pi/L_y ; \quad &n = 0, 1, 2, ... \\
    \Phi_p =& C_p \cos(\lambda_z^p z) \quad &\text{with } \lambda_z^p =& \ p \pi/L_z ; \quad &p = 0, 1, 2, ...
\end{aligned}
\end{equation}

The corresponding eigenfunctions constitute an orthogonal basis. 
To each of these eigenfunctions, we impose the orthonormality constraints:
\begin{equation}
\int_0^{L_x} \Phi_m^2 dx = \int_0^{L_y} \Phi_n^2 dy = \int_0^{L_z} \Phi_p^2 dz = 1
\end{equation}

Evaluating these integrals yields two distinct normalization constants for each of the three coefficients $C_m$, $C_n$ and $C_p$, depending on whether the index is zero or strictly positive. 
A Kronecker delta function combines these cases into a single expression:

\begin{equation}\label{eq:normalisation_constant}
C_m = \sqrt{\frac{2-\delta_{m0}}{L_x}}; \quad C_n = \sqrt{\frac{2-\delta_{n0}}{L_y}}; \quad C_p = \sqrt{\frac{2-\delta_{p0}}{L_z}}
\end{equation}
where $\delta_{i0} = 1$ when $i=0$ and $\delta_{i0} = 0$ otherwise, for $i=m,n,p$.

The normalized one-dimensional (1D) eigenfunctions are therefore:
\begin{equation}
\Phi_{m} =  \sqrt{\frac{2-\delta_{m0}}{L_x}} \cos\left(\frac{m\pi x}{L_x}\right); \quad 
\Phi_{n} =  \sqrt{\frac{2-\delta_{n0}}{L_y}} \cos\left(\frac{n\pi y}{L_y}\right); \quad
\Phi_{p} =  \sqrt{\frac{2-\delta_{p0}}{L_z}} \cos\left(\frac{p\pi z}{L_z}\right)
\end{equation}

Multiplying these 1D orthonormal eigenfunctions gives the 3D orthonormal eigenfunction:
\begin{equation}\label{eq:3Dbasis}
\Phi_{mnp}(\mathbf{x}) = C_m C_n C_p \cos\left(\frac{m\pi x}{L_x}\right)\cos\left(\frac{n\pi y}{L_y}\right) \cos\left(\frac{p\pi z}{L_z}\right)
\end{equation}
with eigenvalues
\begin{equation}
    \lambda_{mnp} = \left( \frac{m\pi}{L_x} \right)^2 + \left( \frac{n\pi}{L_y} \right)^2 + \left( \frac{p\pi}{L_z} \right)^2
\end{equation}

$\Phi_{mnp}$ forms a complete 3D orthonormal basis for the function space of interest and it has the following important and useful properties
\begin{equation}\label{eq:doublephi}
    \int_V \Phi_{mnp} \Phi_{m'n'p'} dV = \delta_{mm'} \delta_{nn'} \delta_{pp'}
\end{equation}
and
\begin{equation}\label{eq:doublegradphi}
    \int_V \nabla \Phi_{mnp} \cdot \nabla \Phi_{m'n'p'} dV = \lambda_{mnp} \delta_{mm'} \delta_{nn'} \delta_{pp'}
\end{equation}
that will allow transforming the Galerkin projection of the heat equation into a system of modal ODEs in time, as shown in the next section.

Based on the spectral theorem for self-adjoint operators \cite{habermanAppliedPartialDifferential2013}, any sufficiently smooth function can be uniquely represented as a convergent series in this basis, for both homogeneous Neumann conditions \eqref{eq:laplaceproblem} or inhomogeneous flux conditions.
Therefore, we make the following ansatz for the temperature field:
\begin{equation}\label{eq:spectral_expansion}
    T(\boldsymbol{x}, t) = \sum_{m'=0}^{\infty} \sum_{n'=0}^{\infty} \sum_{p'=0}^{\infty} \Theta_{m'n'p'}(t) \, \Phi_{m'n'p'}(\boldsymbol{x})
\end{equation}
where the time-dependent modal coefficients $\Theta_{m'n'p'}(t)$ represent the amplitude of each spatial mode.

\subsection{Spectral expansion and Galerkin projection}

The Galerkin projection is used to derive the modal evolution equations for $\Theta_{mnp}$.
The reformulated heat equation~\eqref{eq:Lheatproblem} in $\Omega$ is multiplied by a test function $\Phi_{mnp}$ from the eigenbasis and integrated over the domain $\Omega$:

\begin{equation}\label{eq:galerkin_projection}
\int_{\Omega} \mathcal{L}T \Phi_{mnp} \, dV = \int_\Omega R(T) \Phi_{mnp} \, dV
\end{equation}

Substituting \eqref{eq:spectral_expansion} into this equation, the left hand side is evaluated by applying Green's first identity once to the reference diffusion term and using equations \eqref{eq:doublephi} and \eqref{eq:doublegradphi}:

\begin{equation}\label{eq:lhs_integral}
\begin{aligned}
    &\int_\Omega \mathcal{L}T \, \Phi_{mnp} \, dV \\
    =& \sum_{m'=0}^{\infty} \sum_{n'=0}^{\infty} \sum_{p'=0}^{\infty}
       \left( \overline{\rho}\,\overline{c}\,\dot{\Theta}_{m'n'p'}
       \int_\Omega \Phi_{m'n'p'}\Phi_{mnp}\, dV
       + \overline{k}\,\Theta_{m'n'p'}
       \int_\Omega \nabla\Phi_{m'n'p'}\cdot\nabla\Phi_{mnp}\, dV \right) \\
    & \quad - \overline{k}\oint_{\partial\Omega} \frac{\partial T}{\partial n}\,\Phi_{mnp}\, dS \\
    =& \ \overline{\rho}\,\overline{c}\,\dot{\Theta}_{mnp}
       + \overline{k}\,\lambda_{mnp}\,\Theta_{mnp}
       - \overline{k}\oint_{\partial\Omega} \frac{\partial T}{\partial n}\,\Phi_{mnp}\, dS.
\end{aligned}
\end{equation}

The right hand side of equation~\eqref{eq:galerkin_projection} is evaluated by
applying the divergence theorem to the conductivity-fluctuation term:
\begin{equation}
    \label{eq:forcing_integral}
    \begin{aligned}
        \int_\Omega R(T)\,\Phi_{mnp}\,dV
        =& \ \oint_{\partial\Omega}\left(\widetilde{\boldsymbol{K}}\cdot\nabla T\right)\cdot\boldsymbol{n}\,\Phi_{mnp}\,dS
           - \int_\Omega \nabla\Phi_{mnp}\cdot\widetilde{\boldsymbol{K}}\cdot\nabla T\,dV \\
        & - \int_\Omega \left[ \rho\sum_j L_j \frac{\partial f_j}{\partial T}\dot{T}
           + \left(\overline{\rho}\,\widetilde{c} + \widetilde{\rho}\,\overline{c}
           + \widetilde{\rho}\,\widetilde{c}\right)\dot{T} - S\right]\Phi_{mnp}\,dV
    \end{aligned}
\end{equation}

Equating the two sides, rearranging terms and noting that
\begin{equation}
    \overline{k}\oint_{\partial\Omega}\frac{\partial T}{\partial n}\,\Phi_{mnp}\,dS
    + \oint_{\partial\Omega}\left(\widetilde{\boldsymbol{K}}\cdot\nabla T\right)\cdot\boldsymbol{n}\,\Phi_{mnp}\,dS
    = \oint_{\partial\Omega}\left(\boldsymbol{K}\cdot\nabla T\right)\cdot\boldsymbol{n}\,\Phi_{mnp}\,dS
    = -\oint_{\partial\Omega}\sum_{s=1}^{N_s} q_s\,\Phi_{mnp}\,dS ,
\end{equation} 
yields a neatly system of modal ODEs in time:
\begin{equation}\label{eq:ODE}
\overline{\rho} \,  \overline{c} \;\dot{\Theta}_{mnp} + \overline{k} \, \lambda_{mnp} \,  \Theta_{mnp} = F_{mnp}.
\end{equation}

The forcing
\begin{equation}\label{eq:forcing}
    \begin{aligned}
        F_{mnp} :=& -\int_{\partial\Omega}\sum_{s=1}^{N_s} q_s\,\Phi_{mnp}\,dS
           - \int_\Omega \nabla\Phi_{mnp}\cdot\widetilde{\boldsymbol{K}}\cdot\nabla T\,dV \\
        & - \int_\Omega \left[ \rho\sum_j L_j \frac{\partial f_j}{\partial T}\dot{T}
           + \left(\overline{\rho}\,\widetilde{c} + \widetilde{\rho}\,\overline{c}
           + \widetilde{\rho}\,\widetilde{c}\right)\dot{T} - S\right]\Phi_{mnp}\,dV,
    \end{aligned}
\end{equation} 
gathers all the nonlinear contributions, each of which is temperature dependent.
The forcing is therefore kept as a general, time varying function, evaluated in physical space from the current temperature field, projected back onto the basis, and lagged at the previous iterate within the solver's fixed-point iterative loop.

The linear isotropic reference operator and its associated eigensystem thus reduce the Galerkin projection of the nonlinear heat problem to a system of modal ODEs in time, while retaining all the nonlinearities of the original problem. 
This reduction occurs because the isotropic reference operator has no cross derivative terms coupling the modes; such cross derivatives occur for example in the elliptic static equilibrium equation \cite{moulinecFastNumericalMethod1994a, moulinecNumericalMethodComputing1998, eyreFastNumericalScheme1999}.



\subsection{Time integration scheme}

For the numerical implementation, the exact spectral expansion \eqref{eq:spectral_expansion} is truncated to a finite number of modes $N_x$, $N_y$, $N_z$ per direction, chosen large enough that the resulting truncation error falls below the target accuracy; a problem-specific convergence analysis needs to be performed as shown for the laser scanning example in Section~\ref{sec:results_analytical}.
The system of modal ODEs can be advanced in time using a semi-analytical exponential time differencing scheme. 
The linear reference operator is integrated analytically. 
Nevertheless, since the nonlinear volumetric and surface forcing terms in $F_{mnp}$ in equation~\eqref{eq:ODE} depend on the unknown temperature field at the current time step, an iterative scheme is required.
This work uses the simplest such scheme, the Picard fixed-point iteration; even so, it yields significant computational gains over the optimized FE approach with practically no loss of accuracy.

A first-order exponential time differencing (ETD1) scheme combined with Picard fixed-point iterations is employed. 
The modal decay rate is defined as $\mu_{mnp} = \overline{k}\,\lambda_{mnp} / (\overline{\rho}\,\overline{c})$. 
Within a time step $\Delta t$, the nonlinear forcing $F_{mnp}$ is lagged: held constant across the step and evaluated from the temperature field reconstructed at the previous Picard iterate, $T^{\,n+1,r-1}$. At the Picard iteration $r$ this gives:

\begin{equation}\label{eq:etd1}
\Theta_{mnp}^{\, n+1, r} = E_{mnp} \Theta_{mnp}^{\, n} + Q_{mnp} F_{mnp}^{\, n+1, r-1}
\end{equation}
where $n$ is the time-level index. The exponential operator $E_{mnp}$ and the quadrature coefficient $Q_{mnp}$ are defined as:
\begin{equation}\label{eq:etd1_operators}
E_{mnp} = e^{-\mu_{mnp} \Delta t} \quad \text{and} \quad
Q_{mnp} = \begin{cases}
    (1 - e^{-\mu_{mnp} \Delta t}) / (\mu_{mnp} \overline{\rho} \overline{c}), \quad &\mu_{mnp} \neq 0 \\
    \Delta t/\overline{\rho} \, \overline{c}, \quad &\mu_{mnp} = 0
\end{cases}
\end{equation}

Note that the time integration scheme used here has not been optimized.
Higher-order accurate ETD schemes can be developed without adding significant computational complexity, as the associated higher-order linear operators would remain precomputable and applied as simple mode-wise array multiplications.

Finally, $\Theta_{mnp}(0)$ can be evaluated using the projection of the initial condition \eqref{eq:init_cond} on the orthonormal basis as:
\begin{equation}
    \Theta_{mnp}(0) = \int_\Omega T_{\text{init}} \Phi_{mnp} \, dV
\end{equation}

For a homogeneous initial condition $T_{\text{init}}(\boldsymbol{x}) = \hat{T}$, only the zeroth mode $\Phi_{000} = 1/\sqrt{|\Omega|}$ is non-zero,
which gives $\Theta_{000}(0) = \hat{T} \sqrt{|\Omega|}$ and $\Theta_{mnp}(0) = 0$ for $(m,n,p) \neq (0,0,0)$.

\subsection{Algorithm: SG ETD1 scheme}

Algorithm~\ref{algo:spectral} details the pseudo-code implementation of the SG ETD1 scheme. 
Since $E_{mnp}$ and $Q_{mnp}$ depend only on the modal decay rates $\mu_{mnp}$ and the time step $\Delta t$, they are independent of the nonlinear forcing. 
They are constant for a fixed $\Delta t$ and can be precomputed once before the time loop, leaving only mode-wise multiplications during the Picard iterations.
The nonlinearity of the forcing term stiffens the time integration scheme shown in \eqref{eq:etd1}.
To stabilize the fixed-point iteration, and account for stiffness, a relaxation factor $\omega \in (0,1]$ blends each new modal iterate $\widehat{\Theta}_{mnp}^{\,n+1,r}$ with the previous one.
Its value depends on the stiffness of the problem and is tuned to the largest value for which the Picard iteration still converges monotonically and rapidly.
The iteration is repeated until the relative $L^2$ norm of the fixed-point residual, $\|\widehat{\Theta}_{mnp}^{\,n+1,r} - \Theta_{mnp}^{\,n+1,r-1}\|_{L^2} \,/\, \|\Theta_{mnp}^{\,n+1,r-1}\|_{L^2}$, falls below a tolerance $\epsilon$.
The residual is measured on the un-relaxed update rather than on the difference of successive relaxed iterates, so that the test is independent of $\omega$.

A discrete cosine transform (DCT) operator $\mathcal{D}\{\cdot\}$ maps a discrete spatial field in real space to its eigenfunction space (projection onto the modal basis) to obtain the $(m,n,p)$ modes; $\mathcal{D}^{-1}\{\cdot\}$ performs the inverse operation. 
The DCT along all spatial directions works for heat flux (Neumann) boundary conditions.
For Robin or Dirichlet type boundary conditions, discrete sine transforms may also be needed.

\begin{algorithm}[H]
    \caption{Spectral Galerkin (SG) solver. Each Picard iteration reconstructs the temperature field, projects the nonlinear forcing onto the eigenbasis, and advances every mode by its exact ETD propagator.\label{algo:spectral}}
    \begin{algorithmic}[1]\small
         \STATE \textbf{Pre-compute:} ETD1 propagators $E_{mnp}, Q_{mnp}$ \eqref{eq:etd1_operators} and transform operators $\mathcal{D}, \mathcal{D}^{-1}$. \vspace{4pt}
        \FOR{each time step $n \to n+1$ of size $\Delta t$} \vspace{4pt}
        \STATE $\Theta_{mnp}^{\,n+1,0} \gets \Theta_{mnp}^{\,n}, \quad r \gets 0$ \COMMENT{Start the iteration with the previous modes} \vspace{1pt}
            \REPEAT 
                \STATE $r \gets r+1$ \COMMENT{Picard iteration resolving the nonlinear forcing $F_{mnp}$} \vspace{4pt}
                \STATE $T^{\,n+1,r-1} \gets \mathcal{D}^{-1}\{\Theta_{mnp}^{\,n+1,r-1}\}$ \COMMENT{reconstruct the temperature field} \vspace{4pt}
                \STATE $\dot{T}^{n+1,r-1} = \frac{T^{n+1,r-1} - T^n}{\Delta t}$ \COMMENT{compute the temperature rate} \vspace{4pt}
                \STATE $F_{mnp}^{\,n+1,r-1} \gets \mathcal{D}\{R(T^{\,n+1,r-1})\} - 
                \mathcal{D}_{\partial \Omega}\{R_s(T^{\,n+1,r-1})\}$
                \COMMENT{project the bulk and surface forcing \eqref{eq:Rs}} \vspace{4pt}
                \STATE $\widehat{\Theta}_{mnp}^{\,n+1,r} \gets E_{mnp}\,\Theta_{mnp}^{\,n} + Q_{mnp}\, F_{mnp}^{\,n+1,r-1}$ \COMMENT{ETD1 update \eqref{eq:etd1}} \vspace{4pt}
                \STATE Relax the modes: $\Theta_{mnp}^{\,n+1,r} \gets \Theta_{mnp}^{\,n+1,r-1} + \omega\,\big(\widehat{\Theta}_{mnp}^{\,n+1,r} - \Theta_{mnp}^{\,n+1,r-1}\big)$ \vspace{4pt}
            \UNTIL{$\|\widehat{\Theta}_{mnp}^{\,n+1,r} - \Theta_{mnp}^{\,n+1,r-1}\|_{L^2} \,/\, \|\Theta_{mnp}^{\,n+1,r-1}\|_{L^2} < \epsilon$} \vspace{4pt}
            \STATE $\Theta_{mnp}^{\,n+1} \gets \Theta_{mnp}^{\,n+1,r}$ \COMMENT{converged step} \vspace{4pt}
        \ENDFOR
    \end{algorithmic}
\end{algorithm}

\section{Rapid laser scanning with melting: Simulation setup and numerical implementation}

To demonstrate the capabilities of the proposed SG method, a representative test case of rapid laser scanning on a 316L austenitic stainless steel (henceforth called 316L) substrate is studied. 
This configuration involves different nonlinearities in the form of the latent heat of fusion released and absorbed at the solid--liquid interface, evaporative heat loss at the liquid--gas interface, convective exchange at other boundaries, and temperature-dependent thermophysical properties. 

\subsection{Melting with a moving laser: problem formulation}\label{sec:simsetup_problem}

The cuboid domain $\Omega$ of 316L is taken with directional lengths $L_x = 2$~mm, $L_y = 1$~mm, and $L_z = 0.5$~mm. 
The initial temperature field is $T(x, y, z, 0) = T_0$, where $T_0$ is the ambient temperature.

The governing equation~\eqref{eq:Lheatproblem} takes the following form for this problem:
\begin{equation}\label{eq:laser_problem}
    \underbrace{\overline{\rho} \, \overline{c} \frac{\partial T}{\partial t} - \overline{k} \, \nabla^2 T}_{\mathcal{L}T} = \underbrace{
    \nabla \cdot \left( \widetilde{k} \nabla T \right) - \rho L_{sl} \frac{\partial f_l}{\partial T} \dot{T} - \left( \overline{\rho} \, \widetilde{c} + \widetilde{\rho} \, \overline{c} + \widetilde{\rho} \, \widetilde{c} \right) \dot{T}}_{R(T)}.
\end{equation}
No other source term $S$ is considered.

Since 316L has cubic lattice symmetry in the solid state, its heat conductivity tensor is isotropic such that $\boldsymbol{K}=\left[\overline{k}+\widetilde{k}(T)\right]\mathbb{I}$, where $\widetilde{k}$ is a temperature-dependent scalar.
The reference thermophysical property constants $\overline{\rho}$, $\overline{k}$, and $\overline{c}$ are set to the midpoint of each property's range over the temperature span; this choice follows FFT-based homogenization schemes in mechanics \cite{eyreFastNumericalScheme1999}; those schemes, however, require periodic domains and homogeneous boundary conditions, which the present work does not.
In the solid state, 316L is treated as a single-phase face centered cubic material. 
Solid-state transformations from ferrite to austenite, or the reverse, could occur in 316L stainless steel but they are not treated here for simplicity and without losing any generality of the SG framework or demonstrative capability of the problem.
Phase transition is assumed to occur linearly over the mushy zone between the solidus $T_s$ and the liquidus $T_l$. 
$L_{sl}$ is the specific latent heat of fusion and the liquid fraction $f_l(T)$ is defined as piecewise linear:
\begin{equation}\label{eq:liquid_fraction}
    f_l(T) = 
    \begin{cases} 
        0 & \text{if } T < T_s \\
        \frac{T - T_s}{T_l - T_s} & \text{if } T_s \leq T \leq T_l\\
        1 & \text{if } T > T_l
    \end{cases}
\end{equation}
This approximation captures the continuous release and absorption of latent heat across a diffuse solid--liquid interface \cite{dantzigSolidificationRevisedExpanded2016, vollerEralSourceBasedMethod1991}. 

The Neumann boundary conditions for this problem are summarized as: 

\begin{align}
\underbrace{- \overline{k} \, \boldsymbol{n} \cdot \nabla T}_{\mathcal{L}_s T} =
\underbrace{\begin{cases}
\widetilde{k} \, \boldsymbol{n} \cdot \nabla T + q_{\text{conv}}(T) & z = 0 \\
\widetilde{k} \, \boldsymbol{n} \cdot \nabla T +  q_{\text{evap}}(T) - q_{\text{las}}(x,y,t) & z = L_z \\
\widetilde{k} \, \boldsymbol{n} \cdot \nabla T & \text{otherwise}
\end{cases}}_{R_s(T)}
\end{align}
where on the top surface $z = L_z$, the heat supplied by the laser beam is modeled through $q_{\text{las}}$ and the evaporative heat flux $q_{\text{evap}}$ accounts for the heat loss due to evaporation. 
On the bottom surface $z = 0$, convective heat transfer mimics clamped substrate conditions \cite{weisz-patraultFastSimulationTemperature2020}.
Zero heat flux conditions are imposed on the lateral surfaces.

The expressions of the different surface heat fluxes are given by:
\begin{equation}\label{eq:laser_flux}
  q_{\text{las}}(x,y,t) =
\frac{2 \eta P}{\pi r_b^2}
\exp\!\left[-2\,\frac{(x-x_p(t))^2+(y-y_p(t))^2}{r_b^2}\right]
\end{equation}

\noindent
\begin{equation}\label{eq:evap_flux}
  q_{\text{evap}}(T) = \frac{0.82 \, \Delta H_{lv} \, P_{\text{atm}}}{\sqrt{2 \pi \,R_v \,T}} \exp\!\left[\frac{\Delta H_{lv}}{R_v \, T_{\text{boil}}}\left(1 - \frac{T_{\text{boil}}}{T(x,y,L_z,t)}\right)\right]
\end{equation}

\noindent
\begin{equation} \label{eq:conv_flux}
    q_{\text{conv}}(T)= h_c \left( T - T_0 \right)
\end{equation}
where $P$ is the laser power, $(x_p(t), y_p(t))$ denotes the trajectory of the laser, $\eta$ is absorptivity, $r_b$ is the beam radius, $\Delta H_{lv}$ is the latent heat of vaporization, $P_{\text{atm}}$ is ambient pressure, $R_v$ is the vapor gas constant, $T_{\text{boil}}$ is the boiling temperature, $h_c$ is the baseplate convective coefficient, and $T_0$ is the ambient temperature. 

The expression for $q_{\text{evap}}$ is derived from the Hertz-Knudsen equation coupled with the Clausius-Clapeyron relation, which uses the ambient build chamber pressure ($P_{\text{atm}}$) as its reference state. Furthermore, it incorporates a gas-dynamics correction factor of 0.82 to account for the approximately 18\% of vaporized atoms that condense back onto the surface, as established by Anisimov \cite{anisimovVAPORIZATIONMETALABSORBING1996,knightTheoreticalModelingRapid1979}.

For the validation cases presented in Sections~\ref{sec:results_analytical} and~\ref{sec:results_fe}, the laser follows a single straight pass along the $x$-direction at the domain centreline: $x_p(t) = v_s\,t$, $y_p = L_y/2$, from $x = 0$ to $x = 0.9\,L_x$ (total scan duration $t_{\text{end}} = 12$~ms). 



\begin{figure}[H]
    \centering
    \includegraphics[width=0.5\textwidth]{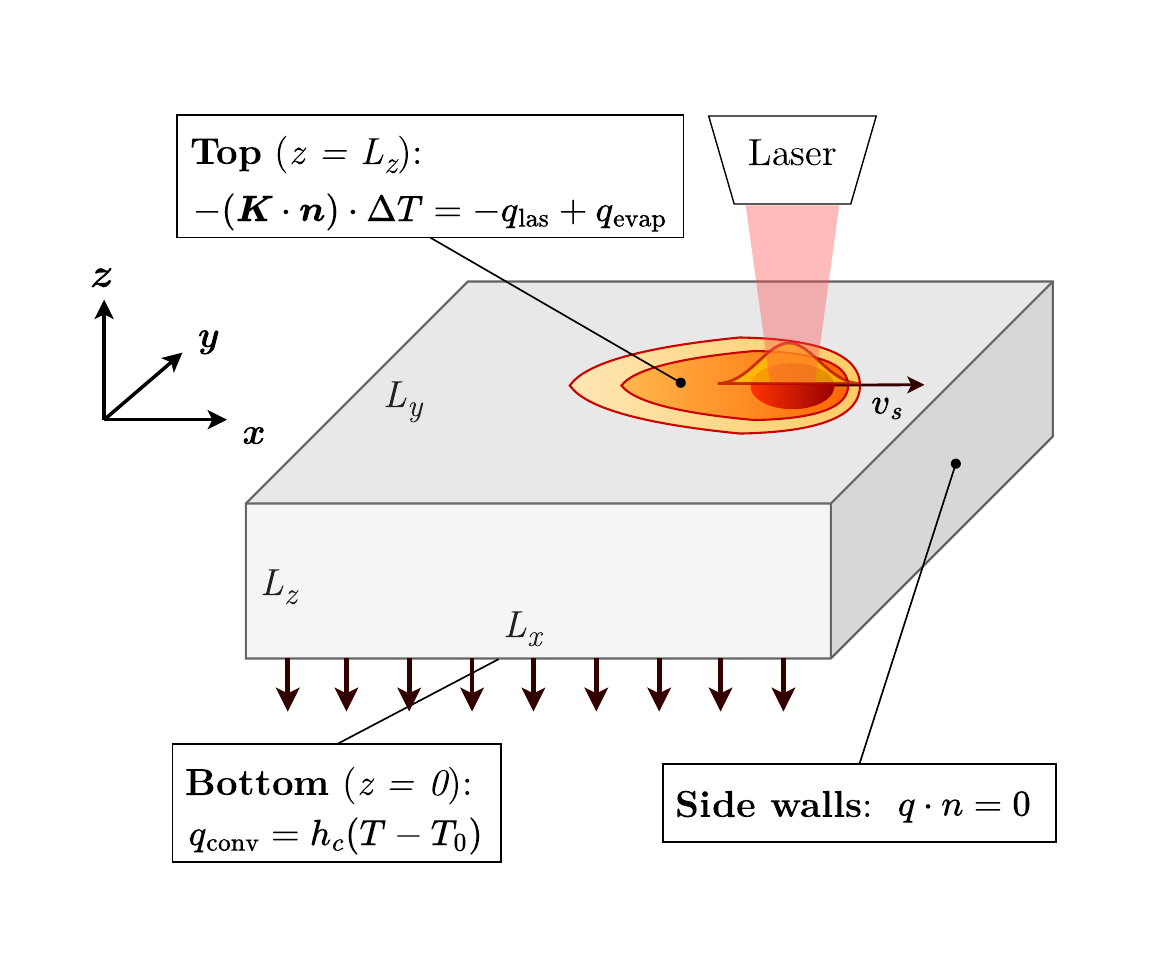}
    \caption{Problem setup: a moving Gaussian laser scans the top surface ($z = L_z$) of the cuboid domain $\Omega = [0,L_x]\times[0,L_y]\times[0,L_z]$ at speed $v_s$, where the absorbed flux $q_{\text{las}}$ competes with the evaporative loss $q_{\text{evap}}$. Convective cooling $q_{\text{conv}} = h_c(T - T_0)$ is imposed on the bottom surface ($z = 0$), and the lateral walls are adiabatic.}
    \label{fig:setup_laser}
\end{figure}

\subsection{Material parameters}

Temperature-dependent thermophysical parameters for 316L stainless steel are used for all numerical simulations performed in this work. 
$\rho$, $k$, and $c_p$ are taken in the temperature-dependent form used by Chadwick et al.~\cite{chadwickMicrostructureDevelopmentLaser2025} (Table~\ref{tab:material_params_Tdep}).
For each property, distinct polynomial expressions are used in the solid ($T \leq T_s$) and liquid ($T \geq T_l$) regions; within the mushy zone, the property is linearly interpolated as shown by equation~\eqref{eq:liquid_fraction}. 
The remaining material and process parameters are listed in Table~\ref{tab:material_params}. 
A temperature-independent variant of the same constants is obtained by evaluating the polynomials of Table~\ref{tab:material_params_Tdep} at $T_0 = 293$~K.

\begin{table}[ht]
\centering
\caption{Temperature-dependent thermophysical properties of 316L stainless steel, after Chadwick et al.~\cite{chadwickMicrostructureDevelopmentLaser2025}. Mushy-zone values are linearly interpolated between solid and liquid using the liquid fraction $f_l(T)$.}
\label{tab:material_params_Tdep}
\begin{tabular}{l l l}
\hline\hline
Property [unit] & Solid ($T \leq T_s$) & Liquid ($T \geq T_l$) \\
\hline
$\rho(T)$ [\si{\kilogram\per\metre\cubed}]
  & $8084.2 - 0.42086\,T$ 
  & $7432.7 + 0.039338\,T$  \\
    &  $-3.8942\times10^{-5}\,T^2$ & $- 1.8007\times10^{-4}\,T^2$
  \\
  \hline 
$c_p(T)$ [\si{\joule\per\kilogram\per\kelvin}]
  & $458.98 + 0.1328\,T$
  & $769.86$ (const.) \\
  \hline 
$k(T)$ [\si{\watt\per\metre\per\kelvin}]
  & $9.248 + 0.01571\,T$
  & $12.41 + 0.003279\,T$ \\
\hline\hline
\end{tabular}
\end{table}

\begin{table}[ht]
\centering
\caption{Constant thermophysical and process parameters used in the simulations (316L SS).}
\label{tab:material_params}
\begin{tabular}{l r l c}
\hline\hline
Parameter & Value & Unit & Reference \\
\hline
\multicolumn{4}{l}{\textit{Material constants}} \\
\hline
Latent heat of fusion ($L_f$) & \num{2.677e5} & \si{\joule\per\kilogram} & \cite{chadwickMicrostructureDevelopmentLaser2025} \\
Latent heat of vaporization ($\Delta H_{LV}$) & \num{7.416e6} & \si{\joule\per\kilogram} & \cite{chadwickMicrostructureDevelopmentLaser2025} \\
Solidus ($T_s$) & \num{1674.15} & \si{\kelvin} & \cite{chadwickMicrostructureDevelopmentLaser2025} \\
Liquidus ($T_l$) & \num{1697.15} & \si{\kelvin} & \cite{chadwickMicrostructureDevelopmentLaser2025} \\
Ambient temperature ($T_0$) & \num{293} & \si{\kelvin} & - \\
Ambient pressure ($P_{\mathrm{atm}}$) & \num{101325} & \si{\pascal} & - \\
Boiling temperature ($T_{\mathrm{boil}}$) & \num{3090} & \si{\kelvin} & \cite{chadwickMicrostructureDevelopmentLaser2025} \\
Vapor gas constant ($R_v$) & \num{150.774} & \si{\joule\per\kilogram\per\kelvin} & \cite{chadwickMicrostructureDevelopmentLaser2025} \\
Absorptivity ($\eta$) & \num{0.30} & {} & \cite{mohananIntergranularStressPlastic2024} \\
Convective coefficient ($h_c$) & \num{3000} & \si{\watt\per\metre\squared\per\kelvin} & \cite{weisz-patraultFastSimulationTemperature2020} \\
\hline
\multicolumn{4}{l}{\textit{Process parameters}} \\
\hline
Laser power ($P$) & \num{24} & \si{\watt} & \cite{chadwickMicrostructureDevelopmentLaser2025} \\
Beam radius ($r_b$) & $30 \times 10^{-6}$ & \si{\metre} & \cite{chadwickMicrostructureDevelopmentLaser2025} \\
Scan speed ($v_s$) & \num{0.15} & \si{\metre\per\second} & \cite{chadwickMicrostructureDevelopmentLaser2025} \\
\hline\hline
\end{tabular}
\end{table}

\subsection{Finite element model setup}

This section validates the SG solver for the nonlinear laser-melting problem against 3D simulations from the FE implementation of the same model (Section~\ref{sec:simsetup_problem}).
Unlike the SG framework, this implementation keeps the full thermophysical properties rather than splitting them into constant reference and fluctuating terms. 
Its weak form follows the same Galerkin projection, without the spectral basis. 
First-order Lagrange elements define the trial and test spaces.
Integrating the diffusion term by parts naturally incorporates the surface fluxes. 
A backward Euler (time-implicit) scheme is used and provides unconditional stability.
The time-discretized weak form is written as:

\begin{equation}
\begin{split}
    \int_{\Omega}\rho c_p\frac{T^{n+1}-T^n}{\Delta t}v\,d\Omega
    +\int_{\Omega}k\nabla T^{n+1}\cdot\nabla v\,d\Omega
    +\int_{\Omega}\rho L_f \dot f_l(T^{n+1},T^n)v\,d\Omega \\
    = \int_{\partial \Omega_{L_z}}\bigl(q_{\mathrm{las}}^{n+1}-q_{\mathrm{evap}}(T^{n+1})\bigr)v\,d\partial \Omega - \int_{\partial \Omega_0}q_{\mathrm{conv}}(T^{n+1})v\,d\partial \Omega
\end{split}
\end{equation}
where the liquid fraction rate $\dot f_l$ is approximated over the time step $\Delta t$ as:

\begin{equation}
\dot f_l= \frac{f(T^{n+1}) - f(T^n)}{\Delta t}
\end{equation}

The FE mesh is refined along the laser trajectory until convergence; a refined element size of \SI{0.6}{\micro\metre} is chosen.
\ref{appendix:impl_fe} details the FE implementation in Algorithm~\ref{alg:implicit_fe_solver} and shows the mesh in Fig.~\ref{fig:mesh_fe}.

\subsection{Numerical implementation}
 
The SG Algorithm~\ref{algo:spectral} applied to the specific case of laser scanning with melting is implemented in Python using both CPU and GPU parallelization. 
Discrete forward and inverse cosine transforms (DCT-II/DCT-III) are used to transition between spectral and physical representations. 
On CPU, the transforms are performed with SciPy's multithreaded scipy.fft.dctn (pocketfft) interface. 
On GPU, since cuFFT provides no native cosine transform, the DCT-II/DCT-III are built directly from CuPy's real FFT (cupy.fft.rfft/cupy.fft.irfft) combined with custom CUDA kernels.
This keeps the implementation fully in Python and allows straightforward CPU/GPU portability.
A convergence analysis is performed for each set of simulation parameters to determine a relaxation factor $\omega$ that ensures monotonic and fast convergence of the Picard iterations.
The Picard iterations are performed until the relative $L^2$ norm of the modal fixed-point residual falls below the tolerance $\varepsilon = 10^{-5}$.

The backward Euler FE scheme of Algorithm~\ref{alg:implicit_fe_solver} is implemented in the open-source software FEniCSx \cite{alnaesFEniCSProjectVersion2015} through its Python interface. 
A Newton-Raphson iterative solver handles all the nonlinearities at each time step. Further implementation details, simulation cases, and code appear in the Appendix and the referenced GitHub repository.

\section{Results}

\subsection{Single-pass laser scanning with melting}

For the laser scanning with melting problem posed in Section~\ref{sec:simsetup_problem}, the SG model is first validated against known analytical and semi-analytical solutions in the fully linear case, i.e., with constant thermophysical properties and no latent heat of fusion, evaporation or convection.
Its predictions are then compared to high-fidelity FE simulations while accounting for all the nonlinearities.


\subsubsection{Linear case: Validation}
\label{sec:results_analytical}

Two classical analytical solutions serve as benchmarks to validate the SG implementation in the linear case. The first, due to Rosenthal \cite{rosenthal1946theory}, gives the steady-state temperature field of a moving point heat source in a semi-infinite medium. 
In a moving frame of reference, where $\xi = x - vt$ denotes the coordinate along the scanning direction, the steady-state temperature field is given by:

\begin{equation}
    T(\xi, y, z) - T_0 = \frac{A P}{2 \pi k R} \exp\left[ -\frac{v (\xi + R)}{2 \alpha} \right]
    \label{eq:rosenthal}
\end{equation}
where $R = \sqrt{\xi^2 + y^2 + z^2}$ is the radial distance from the point source, $\alpha = k / (\rho C_p)$ is the thermal diffusivity, and $v$ is the scanning velocity.  Because the heat input is concentrated at a single point, this solution diverges at the beam center and cannot represent the finite peak temperature of a real distributed laser spot.

The second benchmark, due to Eagar and Tsai \cite{eagar1983temperature}, removes this limitation. Convolving Rosenthal's point source with a Gaussian distribution that matches the physical laser intensity profile yields a more faithful reference for the SG comparison. The resulting semi-analytical expression is:
\begin{equation}
    T(\xi, y, z, t) - T_0 = \frac{A P}{\rho C_p (\pi \alpha)^{1/2}} \int_0^t \frac{1}{\sqrt{\tau} (4 \alpha \tau + r_b^2)} \exp\left[ -\frac{(\xi + v \tau)^2 + y^2}{4 \alpha \tau + r_b^2} - \frac{z^2}{4 \alpha \tau} \right] d\tau
    \label{eq:eagar_tsai}
\end{equation}
where $r_b$ is the characteristic beam radius. 
This equation is numerically integrated using an exponentially decaying time step until convergence is achieved. 
The method of images is used to enforce the zero-flux Neumann boundary conditions on the sides of the cuboid. 
Symmetrical heat source ``images'' are placed around the domain faces; exactly one image source is placed along each principal direction.
This truncated method reproduces the boundary reflections.

\begin{figure}[htbp]
    \centering
    \includegraphics[width=0.55\textwidth]{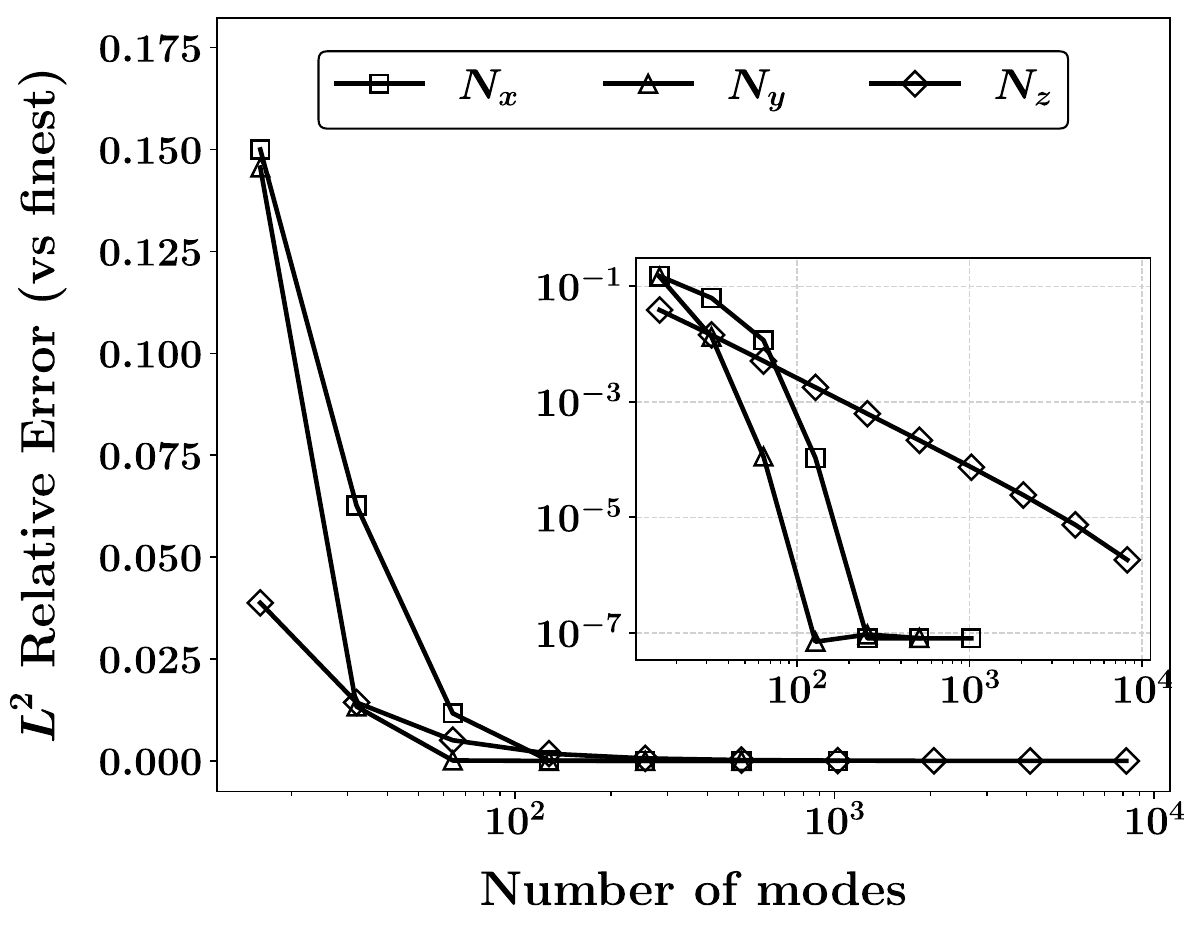}
    \caption{Relative $L^2$ error between the SG solution and the finest-resolution SG reference of each series as a function of the number of modes in each spatial direction, shown on a log-log scale.}
    \label{fig:error_analytical}
\end{figure}

For the SG solver in the linear problem, only the same Gaussian heat source $q_{\text{las}}$ contributes to $F_{mnp}$; all other nonlinear terms in equation~\eqref{eq:forcing} vanish.
The following analysis tests its convergence as a function of the number of modes $N_x$, $N_y$, and $N_z$.
To that end, the $L^2$ norm of the relative difference between each SG prediction and the finest-resolution SG solution is evaluated as the number of modes is varied independently in each spatial direction. 
Figure~\ref{fig:error_analytical} shows this self-convergence: the error decays exponentially in $N_x$ and $N_y$, as expected for smooth in-plane fields, but far more slowly in the vertical direction.
Based on the convergence analysis, the following truncation is used later in this section, for the temperature expansion $N_x = 256$, $N_y = 128$ and $N_z = 512$.

\begin{figure}[H]
    \centering
    \includegraphics[width=0.6\textwidth]{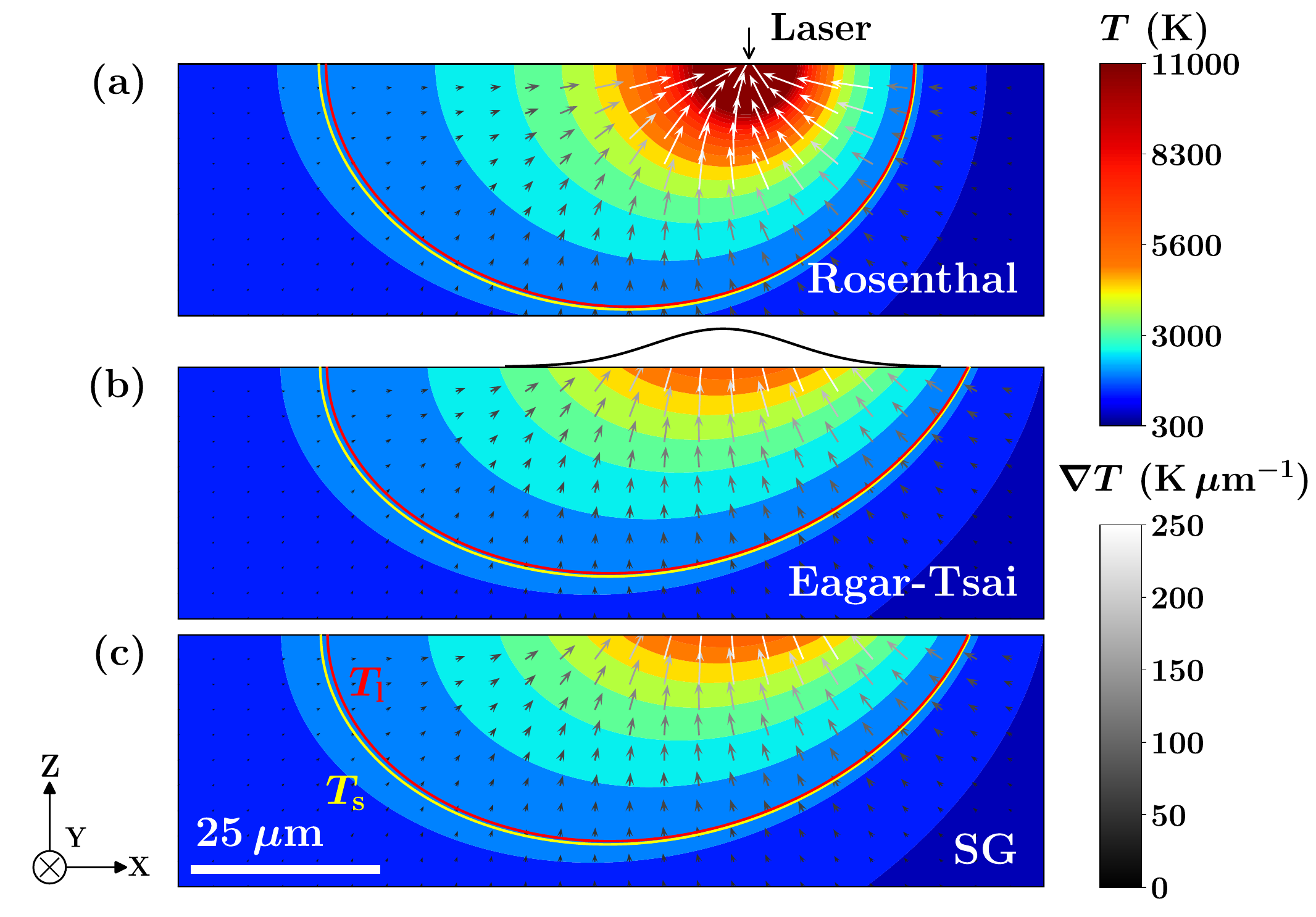}
    \caption{Contour plots of temperature predicted for the linear case with (a) the Rosenthal analytical solution, (b) the Eagar-Tsai semi-analytical solution, and (c) the SG model on the cross-sectional $xz$-plane. The yellow and red isolines mark the solidus ($T_s$) and liquidus ($T_l$). The length and inclination of the greyscale arrows indicate the in-plane magnitude and direction of the temperature gradients.}
    \label{fig:cut_views_linear_y}
\end{figure}

\begin{figure}[H]
    \centering
    \begin{subfigure}[b]{0.50\textwidth}
        \centering
        \includegraphics[width=\textwidth]{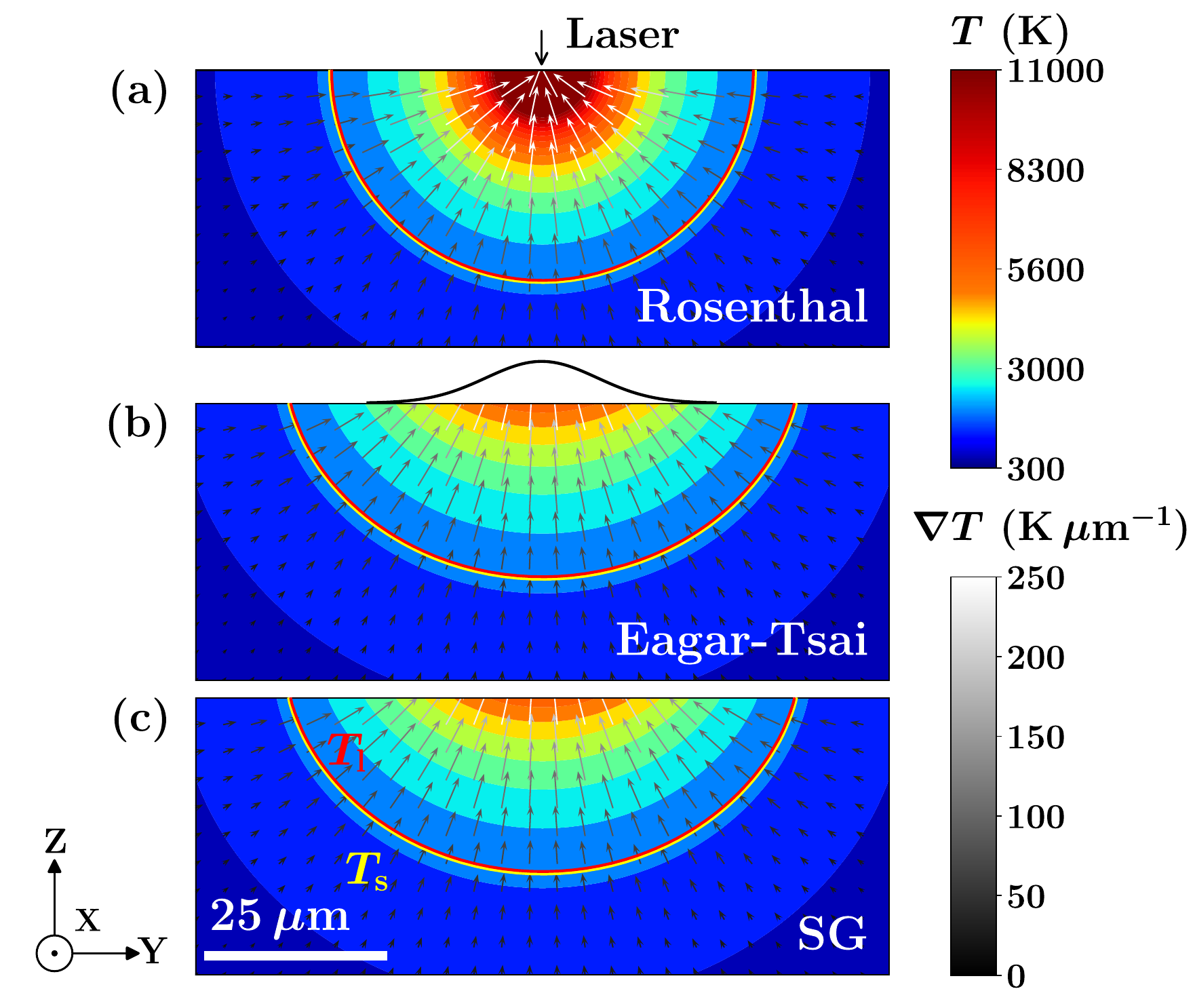}
    \end{subfigure}\hfill
    \begin{subfigure}[b]{0.42\textwidth}
        \centering
        \includegraphics[width=\textwidth]{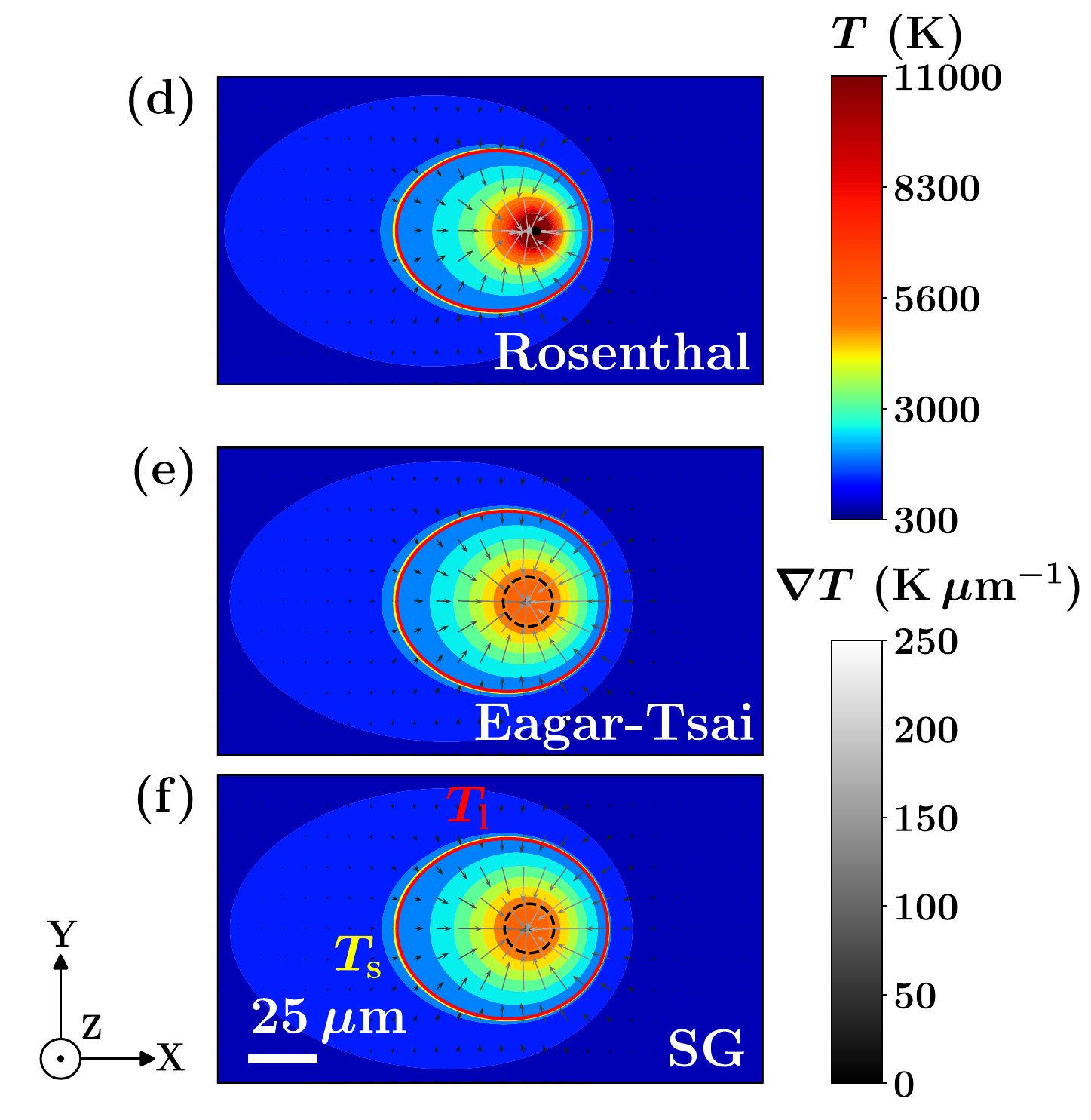}
    \end{subfigure}
    \caption{Contour plots of temperature predicted for the linear case with (a, d) the Rosenthal analytical solution, (b, e) the Eagar-Tsai semi-analytical solution, and (c, f) the SG model on the cross-sectional (a, b, c) $yz$-plane and (d, e, f) $xy$-plane where the dot and dotted lines show the laser source. The yellow and red isolines mark the solidus ($T_s$) and liquidus ($T_l$). The length and inclination of the greyscale arrows indicate the in-plane magnitude and direction of the temperature gradients.}
    \label{fig:cut_views_linear_xz}
\end{figure}

\begin{figure}[H]
    \centering
    \includegraphics[width=0.78\textwidth]{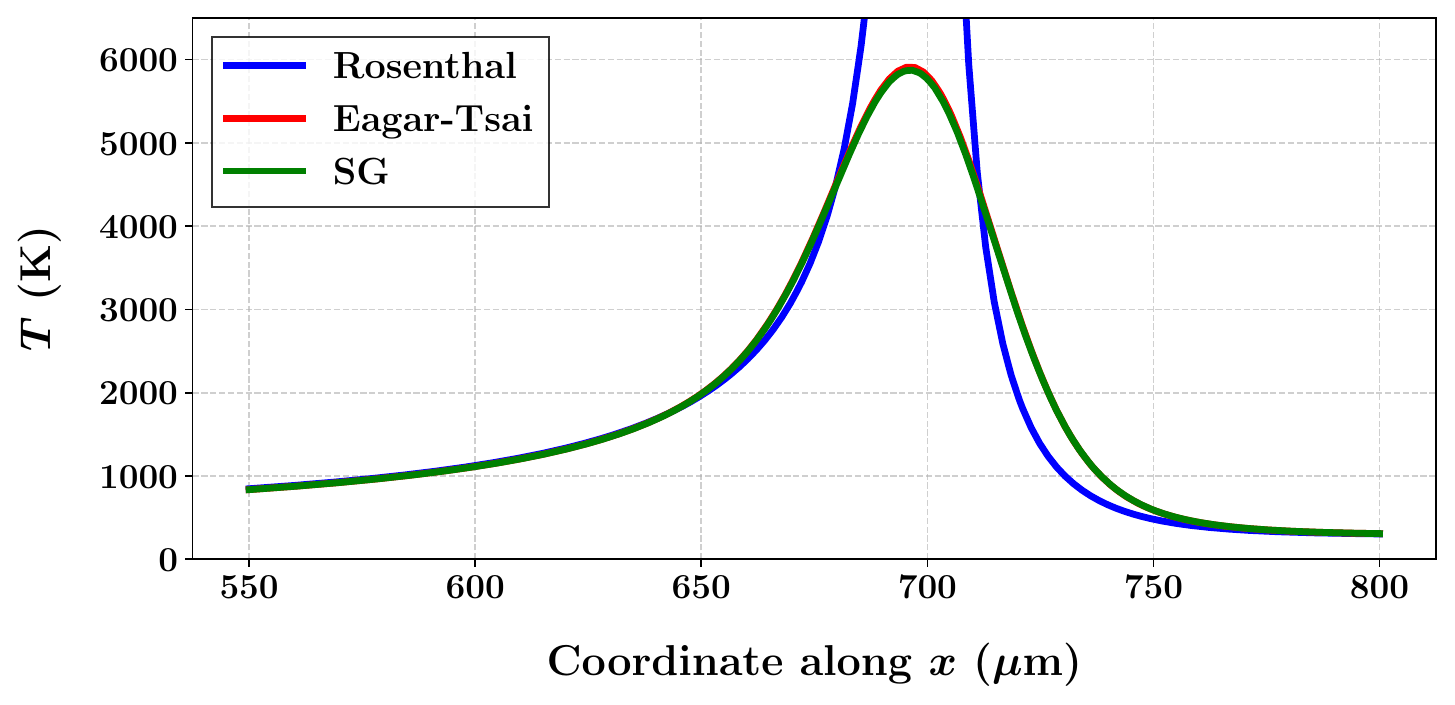}
    \caption{Temperature distribution along the centerline on the top surface of the domain, comparing the Rosenthal, Eagar-Tsai, and SG predictions for the linear heat problem, and demonstrating the excellent agreement between the SG and the Eagar-Tsai solutions.}
    \label{fig:linear_lines}
\end{figure}

Figures~\ref{fig:cut_views_linear_y} and~\ref{fig:cut_views_linear_xz} compare the predictions of the SG simulations at steady state against the Rosenthal and Eagar-Tsai solutions.
All three models have an excellent match in the far field (far from the heat source).
While the Rosenthal solution models only a point heat source, the Eagar-Tsai and SG formulations account for the finite radius of the Gaussian beam and reduce the maximum temperature.
In general, an excellent agreement is obtained between the Eagar-Tsai and SG predictions.
Figure~\ref{fig:linear_lines} further demonstrates this match by plotting the temperature profile along the centerline on the top surface ($z=L_z$) of the domain. 
The relative $L^2$ difference between the SG and Eagar-Tsai predictions is less than 0.1\%.
This comparison validates the reference homogeneous problem, which is the core of the SG model.

\subsubsection{Nonlinear case: Comparison with FE simulations}
\label{sec:results_fe}

\begin{figure}[htbp]
    \centering
    \includegraphics[width=0.55\textwidth]{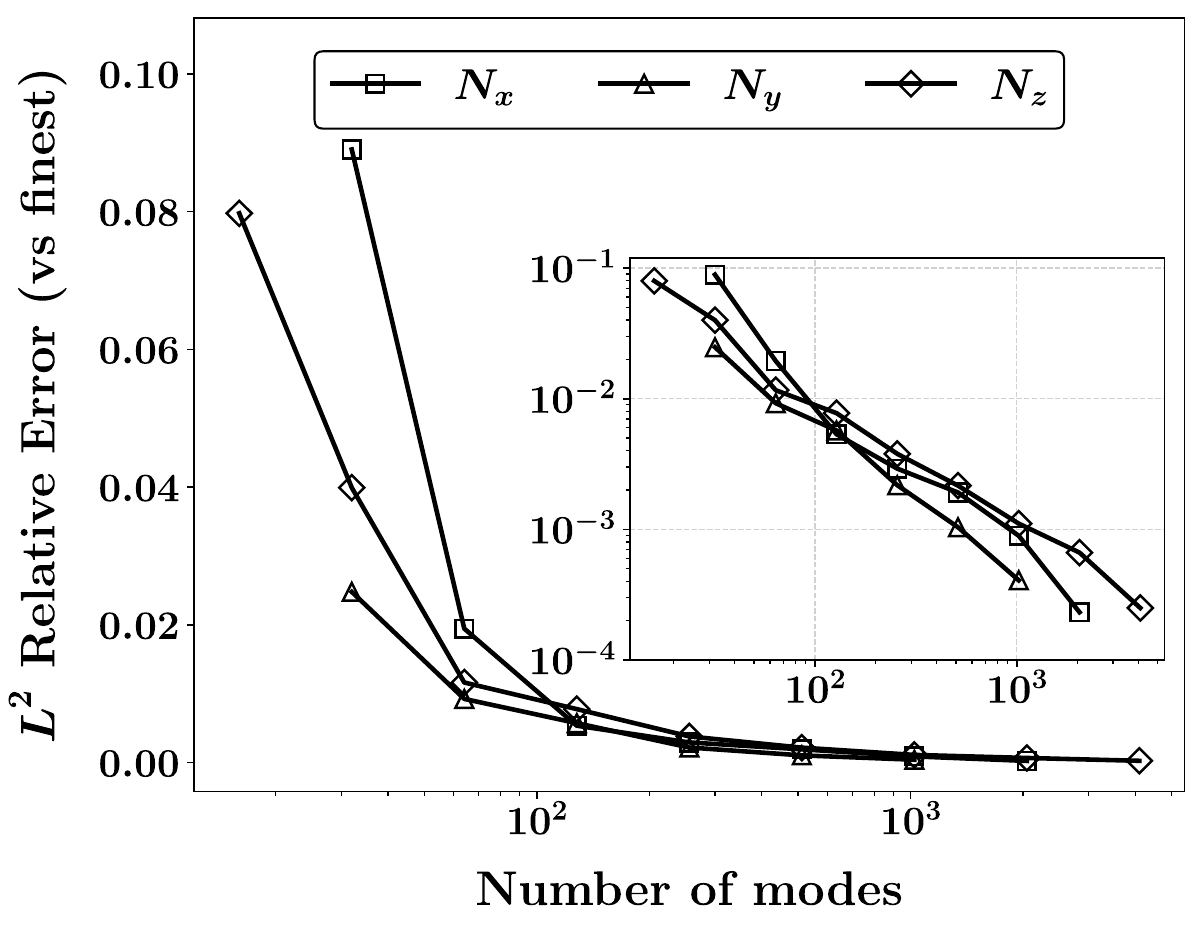}
    \caption{Relative $L^2$ error between the SG and the FE solutions measured against the number of modes in each spatial direction and shown on a log-log scale.}
    \label{fig:error_FE_axes}
\end{figure}

Following the validation of the linear core of the SG method i.e., the homogeneous reference problem and modal integration, the laser-scanning simulation is repeated in the fully nonlinear regime with the temperature-dependent material properties, latent heat of fusion, and evaporative cooling. 
Since closed-form solutions do not exist in the nonlinear case, the SG predictions are benchmarked against the high-fidelity FE simulations using the same simulation parameters. 
This comparison serves a twofold purpose: to confirm that the SG solver retains its accuracy with the nonlinear physics, and to verify that it converges towards a numerical reference.
A convergence analysis is performed for the SG solver in this fully nonlinear regime as a function of $N_x$, $N_y$ and $N_z$. 
Fig. \ref{fig:error_FE_axes} shows the result of this analysis. 
Based on this analysis, a truncation of $N_x = 256$, $N_y = 128$, and $N_z = 768$ are used for comparison with the FE predictions as a tradeoff between accuracy and computational time. <

Convergence is slower than in the linear case (Fig.~\ref{fig:error_analytical}), and slowest in the vertical direction, where the homogeneous-Neumann basis represents the surface flux only weakly and needs many modes to resolve the surface temperature. 
The error nevertheless drops with further refinement, and any target accuracy remains attainable; this motivates the higher vertical resolution $N_z = 768$ used here. 
Section~\ref{sec:discussion} identifies the mechanism that fixes direction-dependent convergence speed and outlines a strategy to accelerate the vertical convergence.

\begin{figure}[H]
    \centering
    \includegraphics[width=0.6\textwidth]{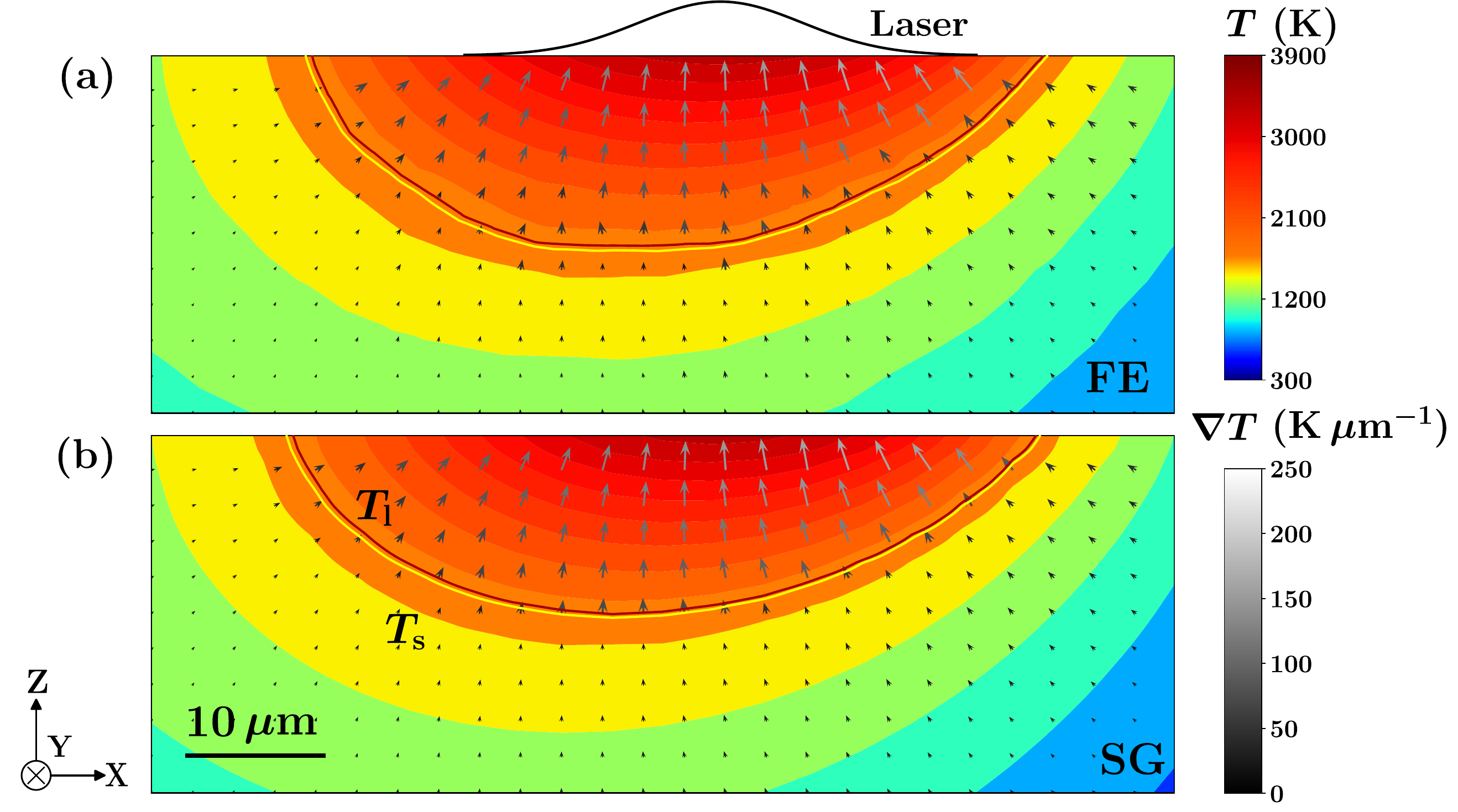}
    \caption{Contour plots of temperature predicted for the nonlinear case with (a) the FE and (b) the SG solvers on the cross-sectional $xz$-plane. The yellow and red isolines mark the solidus ($T_s$) and liquidus ($T_l$). The length and inclination of the greyscale arrows indicate the in-plane magnitude and direction of the temperature gradients.}
    \label{fig:cut_views_FE_SG_y}
\end{figure}

\begin{figure}[H]
    \centering
    \begin{subfigure}[b]{0.60\textwidth}
        \centering
        \includegraphics[width=\textwidth]{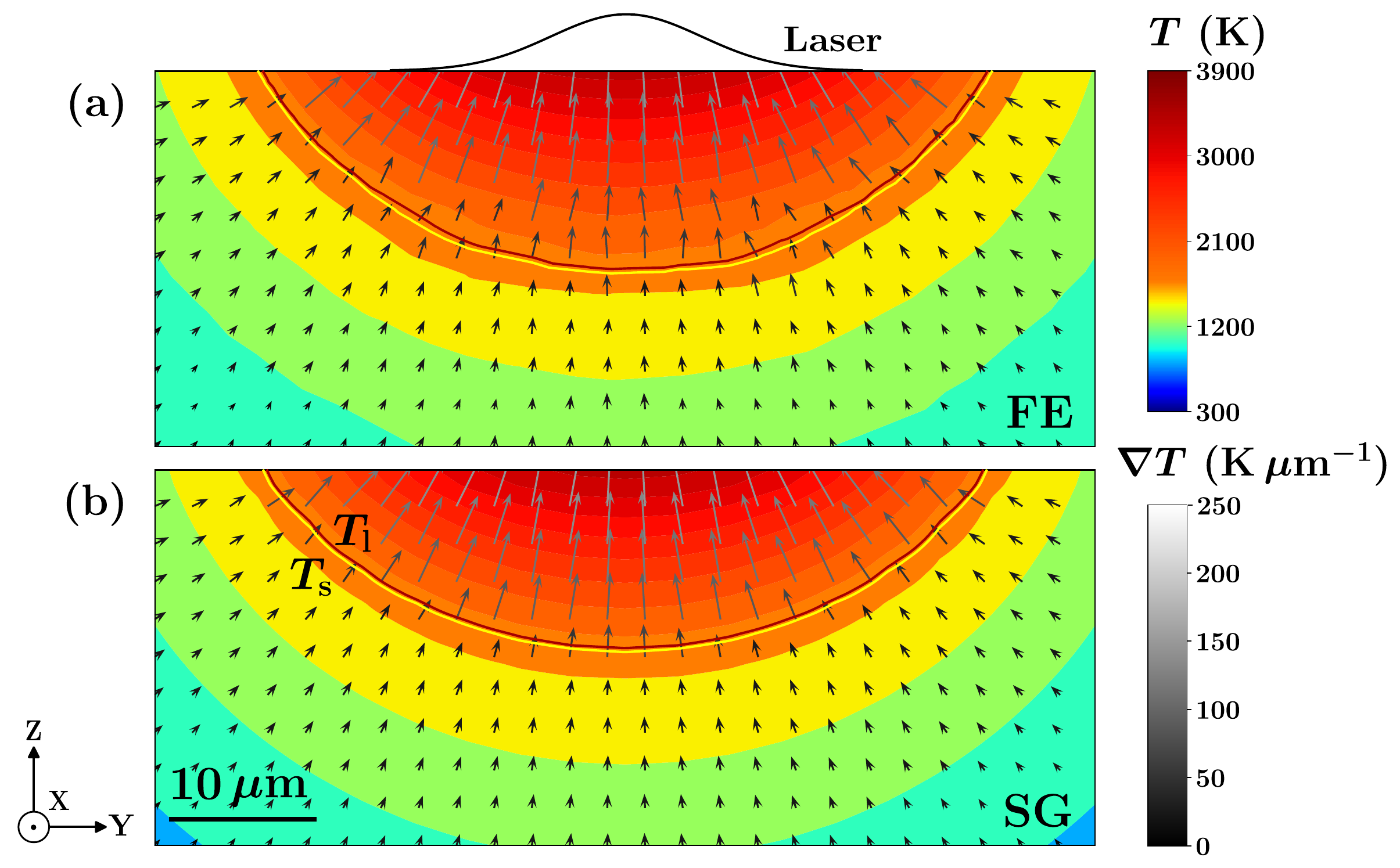}
    \end{subfigure}\hfill
    \begin{subfigure}[b]{0.38\textwidth}
        \centering
        \includegraphics[width=\textwidth]{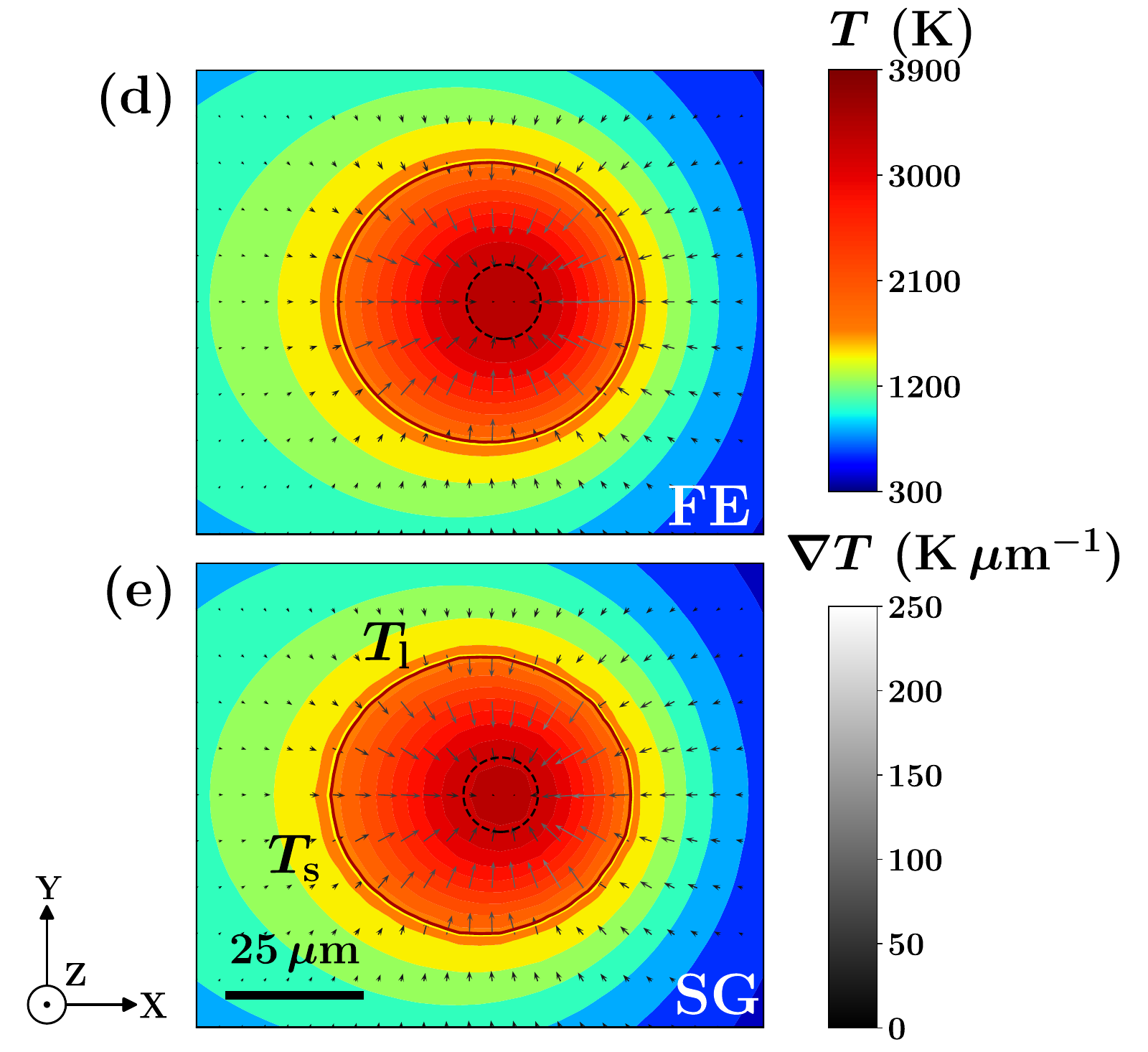}
    \end{subfigure}
    \caption{Contour plots of temperature predicted for the nonlinear case with (a, d) the FE model and (b, e) the SG model on the cross-sectional (a, b) $yz$-plane and (d, e) $xy$-plane where the dotted lines show the laser source. The yellow and red isolines mark the solidus ($T_s$) and liquidus ($T_l$). The length and inclination of the greyscale arrows indicate the in-plane magnitude and direction of the temperature gradients.}
    \label{fig:cut_views_FE_SG_xz}
\end{figure}

\begin{figure}[H]
    \centering
    \includegraphics[width=0.78\textwidth]{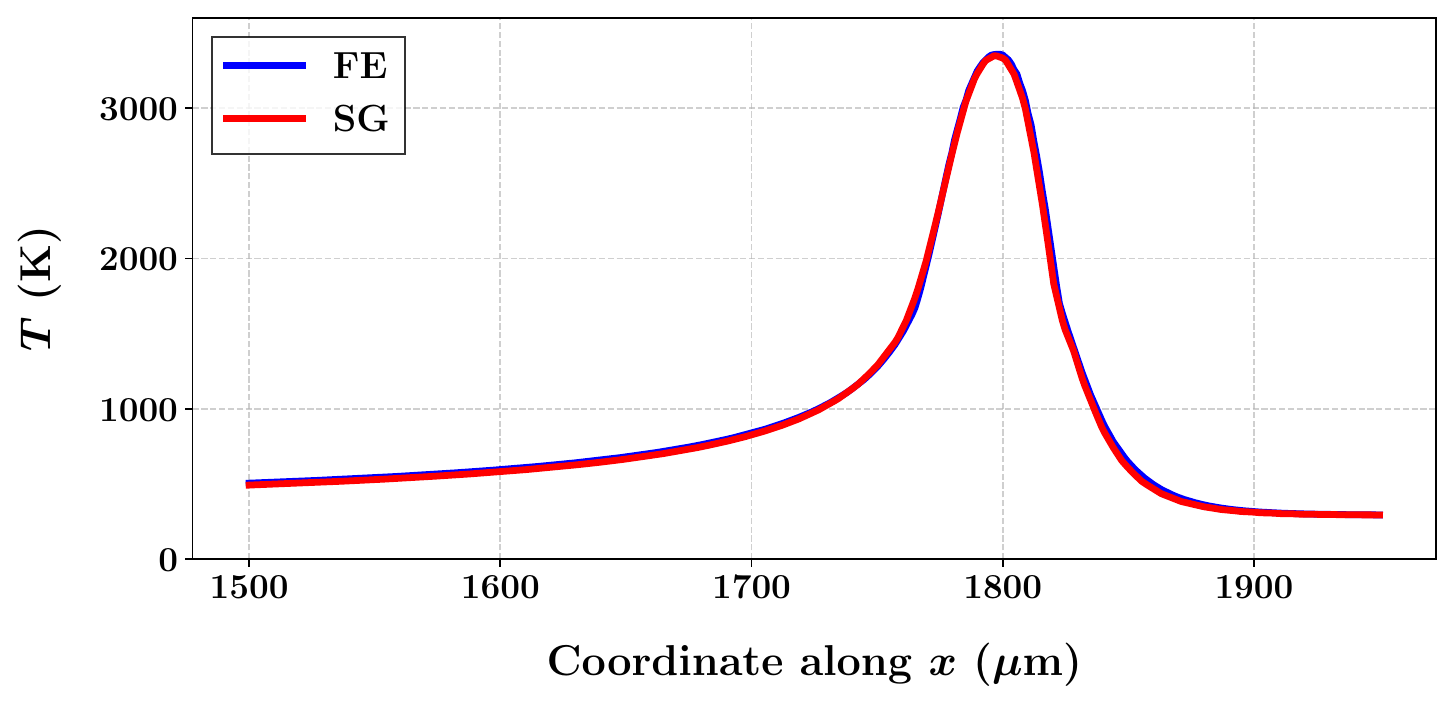}
    \caption{Temperature distribution along the centerline on the top surface of the domain, comparing the FE and SG predictions for the nonlinear heat problem, and demonstrating the excellent agreement between the SG and the FE solutions.}
    \label{fig:nonlinear_lines}
\end{figure}

Figs.~\ref{fig:cut_views_FE_SG_y}
and~\ref{fig:cut_views_FE_SG_xz} present cross-sectional views of the temperature
distribution predicted by the SG and FE models once steady-state is reached. 
The predictions are in excellent agreement: the SG predicted peak temperature differs only by \SI{0.93}{\percent} (\SI{31.3}{\kelvin}). 
The SG solver reproduces with high accuracy both the steep thermal gradients near the heat source and the shape of the melt pool. Figure~\ref{fig:nonlinear_lines} confirms this through the
temperature profile along the centerline on the top surface of the domain, where the SG and FE solutions are nearly indistinguishable.
The relative $L^2$ difference between the SG and FE predictions is \SI{0.67}{\percent} further confirming the excellent agreement.

\subsubsection{Complexity analysis and performance comparison}


The SG solver's computational cost is governed by the spectral transforms mapping the temperature field from modal to physical space at every step. With temperature-dependent properties, the property correction must be evaluated globally, so these transforms dominate the per-step cost and set the overall complexity to $\mathcal{O}(N_{\text{vol}} \log N_{\text{vol}})$ in the total number of degrees of freedom.

\begin{figure}[htbp]
    \centering
    \begin{subfigure}[b]{0.45\textwidth}
        \centering
        \begin{tikzpicture}
            \node[anchor=south west, inner sep=0] (image) at (0,0) {\includegraphics[width=\textwidth]{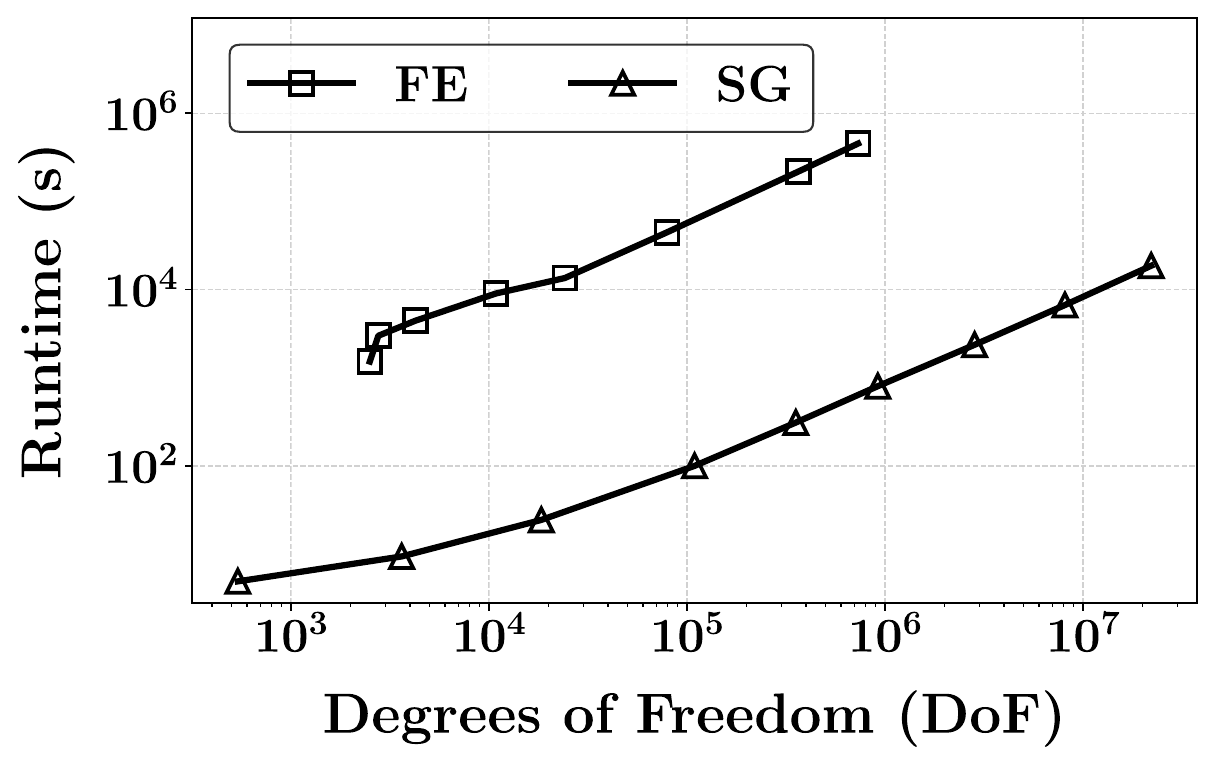}};
            \begin{scope}[x={(image.south east)},y={(image.north west)}]
                \node[anchor=south west, fill=white, fill opacity=0.8, text opacity=1, rounded corners=2pt, inner sep=3pt] at (0.02, 0.02) {\textbf{(a)}};
            \end{scope}
        \end{tikzpicture}
        \phantomcaption
        \label{fig:runtime_vs_dof}
    \end{subfigure}\hfill
    \begin{subfigure}[b]{0.45\textwidth}
        \centering
        \begin{tikzpicture}
            \node[anchor=south west, inner sep=0] (image) at (0,0) {\includegraphics[width=\textwidth]{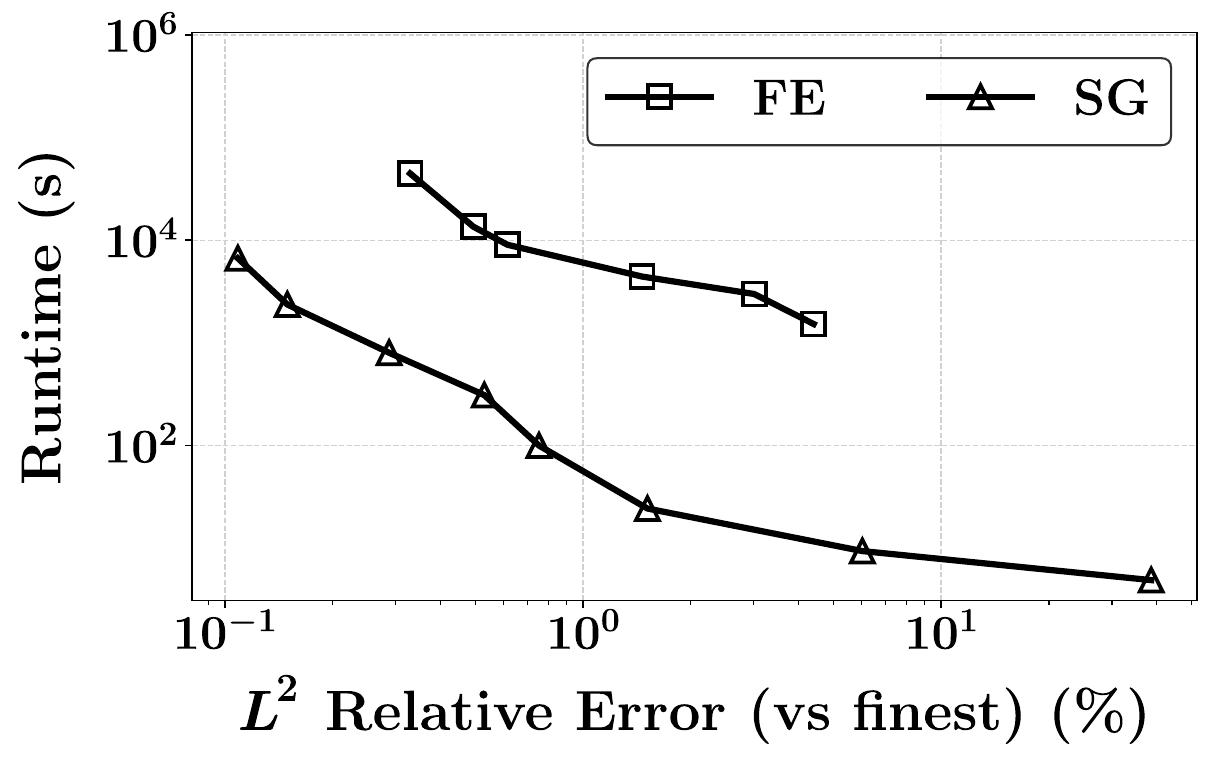}};
            \begin{scope}[x={(image.south east)},y={(image.north west)}]
                \node[anchor=south west, fill=white, fill opacity=0.8, text opacity=1, rounded corners=2pt, inner sep=3pt] at (0.02, 0.02) {\textbf{(b)}};
            \end{scope}
        \end{tikzpicture}
        \phantomcaption
        \label{fig:runtime_vs_error}
    \end{subfigure}
    \caption{Performance comparison between the SG and FE methods. (a) Total execution time plotted against the number of degrees of freedom. (b) Work-precision diagram displaying the runtime as a function of the numerical error.}
\end{figure}

As an initial benchmark for complexity, the SG method is compared against the FE implementation for the same problem, while ensuring the hardware allocation is the same. 
Both codes were executed on a single core of a Ryzen 9 5900X CPU. 
Figure~\ref{fig:runtime_vs_dof} displays the total execution time plotted against the number of degrees of freedom and 
Figure~\ref{fig:runtime_vs_error} presents a work-precision diagram, illustrating the runtime as a function of the $L^2$ relative error. 
The CPU implementation of the SG solver has two orders of magnitude greater computational efficiency than the FE approach, while consistently incurring two orders of magnitude lower relative error.

\begin{figure}[htbp]
    \centering
    \includegraphics[width=0.45\textwidth]{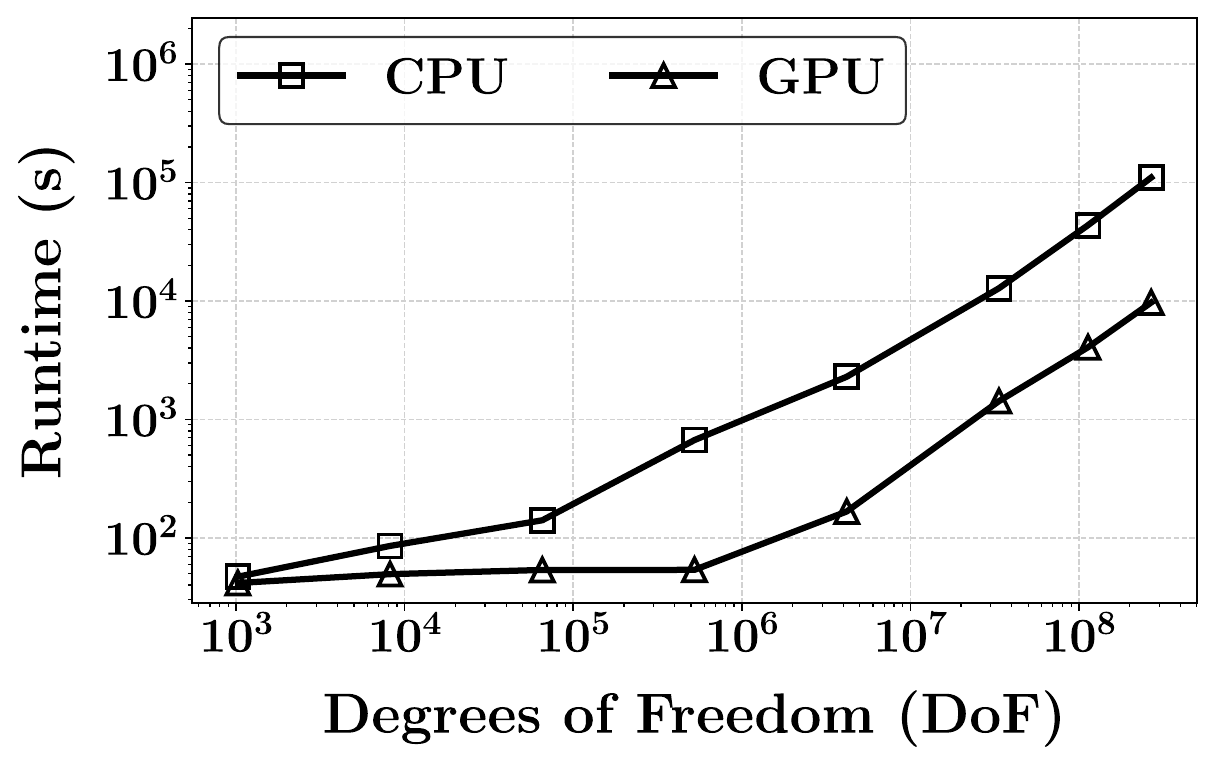}
    \caption{Computational complexity analysis of the SG solver showing CPU and GPU runtimes for varying grid sizes. Both curves indicate asymptotic quasi-linear behavior.}
    \label{fig:complexity_analysis}
\end{figure}

We also assess how the algorithm scales by running the SG solver at increasing grid resolutions on both processors, and recording the wall-clock runtime on a NVIDIA Quadro RTX 5000 GPU (16 GB) and on a dual Intel Xeon Gold 5220R CPU (48 cores / 96 threads).
As Fig.~\ref{fig:complexity_analysis} shows, the GPU implementation runs faster than the CPU one by an order of magnitude.
On coarse grids, a fixed cost per time step dominates the runtime. This cost is due to launch time and data transfer. The algorithmic complexity is visible when the grid becomes large enough, near $N_{\text{vol}} = 5.0\times10^{5}$ on the CPU and $N_{\text{vol}} = 5.0\times10^{6}$ on the GPU. Beyond these resolutions both curves follow a slope close to $1$ on the logarithmic scale, which matches the asymptotic $\mathcal{O}(N_{\text{vol}} \log N_{\text{vol}})$ complexity.

For the cases presented in section \ref{sec:results_fe}, 
with the mesh refined to \SI{0.6}{\micro\metre} in the melt-pool region (2,586,826 degrees of freedom), 
the FE solver required approximately 189 hours on a single core of the Ryzen~9~5900X.
With the same hardware and identical simulation parameters, the SG algorithm resolved a $256\times128\times768$ grid (25,165,824 degrees of freedom) in 30.6 hours, $6.2\times$ faster despite an order of magnitude more degrees of freedom. 
Running the same computation on a full NVIDIA Quadro~RTX~5000 GPU reduced the runtime to 0.83 hours, 
a $227\times$ speed-up relative to the single-core FE baseline.  
These results show that, with the same hardware, the SG method significantly outperforms FE even while resolving more degrees of freedom.
Furthermore, its regular-grid structure is highly suited for an efficient GPU implementation, gaining a further one order of magnitude in speedup. 
A broader discussion on these numbers is presented in Section~\ref{sec:discussion}.

\subsection{Revisiting an existing part-scale multi-pass laser scan-strategy}

The previous section presented a rigorous verification and benchmarking of the SG method against linear analytical and semi-analytical solutions, and nonlinear FE implementations. 
In this section, we utilize the significant computational speedup and high accuracy of the SG method to demonstrate significant improvement in predictions of a part-scale multipass surface scanning strategy at a high resolution that is impractical to achieve with FE simulations.
The SG method is applied to the laser scan-strategy optimization study of Ramani et al.~\cite{ramaniSmartScanIntelligentScanning2022} for the powder bed fusion (PBF) of a 316L stainless steel substrate. Ramani et al. proposed a thermal uniformity metric as a critical factor in PBF processes associated with residual stresses, delamination, and out-of-plane deformation, particularly when processing thin plates. 
This metric is defined as
\begin{equation}\label{uniformity_metric}
R(t) = \frac{1}{N_{\text{grid}} T_m} \sqrt{ \sum_{x,y,z} \left( T(x,y,z, t) - T_{\text{avg}}(t) \right)^2 },
\end{equation}
where $N_{\text{grid}}$ is the number of measurement points, $T_m$ is the melting temperature, and $T_{\text{avg}}(t)$ is the domain's average temperature. This metric is proportional to the normalized standard deviation of the temperature field within the processed component.

To compute this metric, Ramani et al.~employed a finite difference model reduced via radial basis functions~\cite{ramaniSmartScanIntelligentScanning2022} as a baseline. 
Their case study compares four different scan patterns on a $50 \times 50$\SI{}{\milli\metre^2} region on the top wide surface of a $60 \times 60 \times 1$~\SI{}{\milli\metre^3} thick 316L plate.
They divide the scanned region into $100$ square islands, numbered  $1, 2, 3, \ldots, 100$ so that neighbouring islands follow in numbering.
The laser scans each island with $25$ hatch lines spaced \SI{200}{\micro\metre} apart and reverses its direction between adjacent lines and the hatch direction rotates by \SI{90}{\degree} from one island to the next.
Four scan strategies are employed: (i) The successive strategy follows an island numbering, $1, 2, 3, \ldots, 100$, so the laser always move from an island to its neighbour. (ii) The successive chessboard strategy first scans the odd-numbered islands, $1, 3, 5, \ldots, 99$, and then returns to fill the even-numbered ones. (iii) The least heat influence (LHI) strategy places each new island as far as possible from those already scanned~\cite{kruthSelectiveLaserMelting2004} to maximize the distance between the next island and all previously scanned ones. (iv) Finally, SmartScan is a greedy optimization approach: at each step, a reduced-order thermal model predicts the temperature field that each candidate island would create, and the strategy selects the island minimizing the resulting value of $R$.

The SG method is used to simulate these four scan strategies using identical processing parameters. 
The number of modes is set to match the reference grid in the $x$ and $y$ directions; however, a finer resolution is employed in the $z$ direction to preserve the accuracy of the SG method. 
The simulation parameters are summarized in Table~\ref{tab:ramani_params}.

\begin{table}[ht!]
\centering
\caption{Process parameters and material properties used for the PBF case study from Ramani et al.~\cite{ramaniSmartScanIntelligentScanning2022}.}
\label{tab:ramani_params}
\begin{tabular}{l r l}
\hline\hline
Parameter & Value & Unit \\
\hline
Laser power ($P$) & \num{200} & \si{\watt} \\
Spot diameter ($2R_b$) & \num{77} & \si{\micro\metre} \\
Absorptance ($\lambda$) & \num{0.37} & {} \\
Scan speed ($v_s$) & \num{600} & \si{\milli\metre\per\second} \\
Hatch spacing & \num{200} & \si{\micro\metre} \\
Thermal conductivity ($k_t$) & \num{22.5} & \si{\watt\per\metre\per\kelvin} \\
Thermal diffusivity ($\alpha$) & \num{5.632e-6} & \si{\metre\squared\per\second} \\
Melting temperature ($T_m$) & \num{1658} & \si{\kelvin} \\
Convective coefficient ($h$) & \num{25} & \si{\watt\per\metre\squared\per\kelvin} \\
Ambient temperature ($T_a$) & \num{293} & \si{\kelvin} \\
Initial temperature ($T(x,y,z,0)$) & \num{293} & \si{\kelvin} \\
\hline\hline
\end{tabular}
\end{table}

The SG implementation is first validated against the same thermal problem solved by Ramani et al. \cite{ramaniSmartScanIntelligentScanning2022} that only accounts for the Gaussian laser but neglects all other nonlinear effects. 
To ensure a rigorous comparison, the thermal uniformity metric is evaluated precisely at the radial basis function collocation points defined in \cite{ramaniSmartScanIntelligentScanning2022}. 

\begin{figure}[ht!]
    \centering
    \begin{subfigure}[b]{0.46\textwidth}
        \centering
        \includegraphics[width=\textwidth]{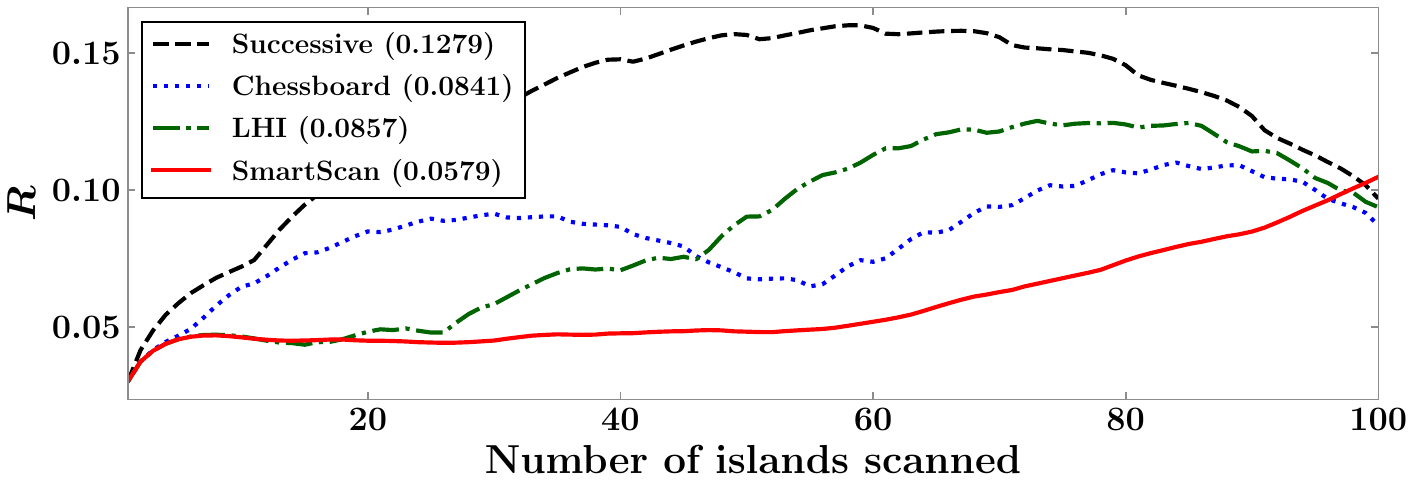}
        \caption{SG model}
        \label{fig:ramani_rep_a}
    \end{subfigure}
    \hfill
    \begin{subfigure}[b]{0.47\textwidth}
        \centering
        \includegraphics[width=\textwidth]{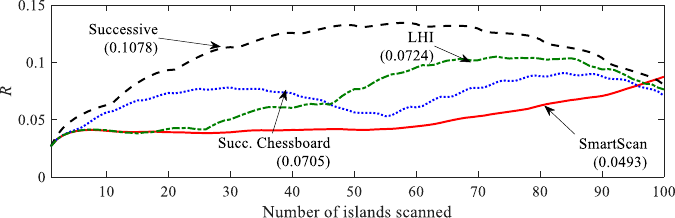}
        \caption{Ramani et al. \cite{ramaniSmartScanIntelligentScanning2022}}
        \label{fig:ramani_rep_b}
    \end{subfigure}
    \vskip\baselineskip
    \begin{subfigure}[b]{0.47\textwidth}
        \centering
        \includegraphics[width=\textwidth]{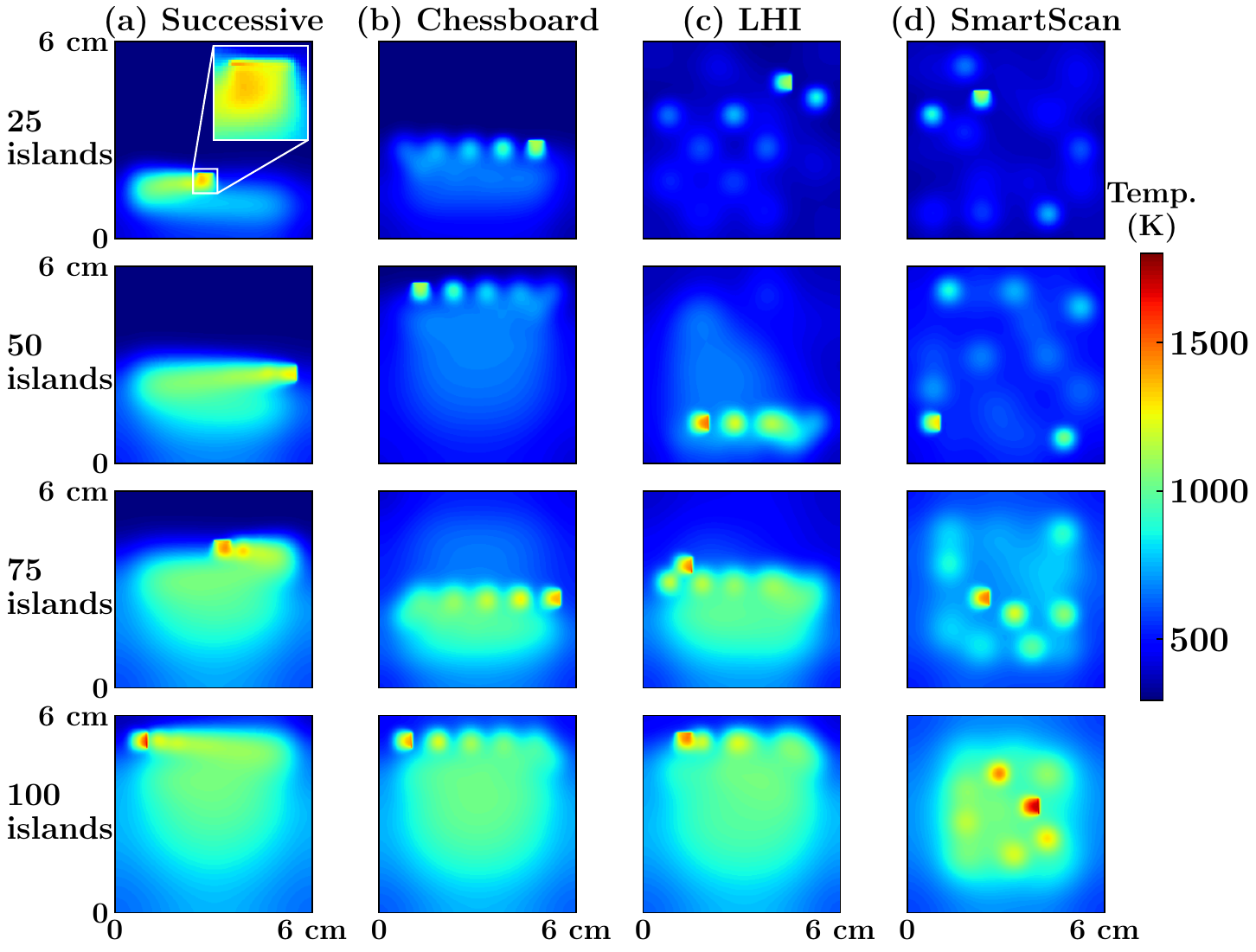}
        \caption{SG model}
        \label{fig:ramani_rep_c}
    \end{subfigure}
    \hfill
    \begin{subfigure}[b]{0.51\textwidth}
        \centering
        \includegraphics[width=\textwidth]{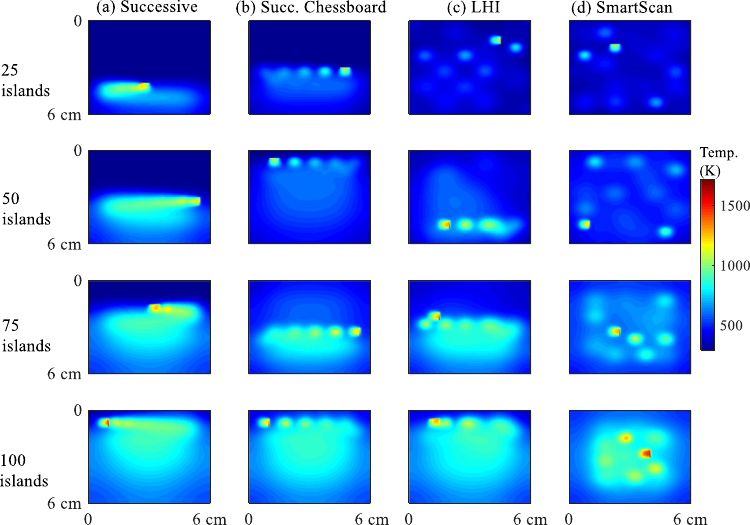}
        \caption{Ramani et al. \cite{ramaniSmartScanIntelligentScanning2022}}
        \label{fig:ramani_rep_d}
    \end{subfigure}
    \caption{Thermal uniformity metric $R(t)$ for the different scan strategies using the (a) SG model and showing good agreement with (b) the results of Ramani et al.~\cite{ramaniSmartScanIntelligentScanning2022}. Temperature field at the end of the scan for the different strategies using the (c) SG model vs. (d) the results of Ramani et al.~\cite{ramaniSmartScanIntelligentScanning2022}; the inset in (c) gives a closer view of one of the islands. The original figures~\cite{ramaniSmartScanIntelligentScanning2022} have been reproduced here with permission from Elsevier.}
    \label{fig:ramani_comparison}
\end{figure}

The SG model completed this linear simulation in \SI{36.7}{\minute}. 
Reproducing the published thermal history necessitated a domain height of $200\,\si{\micro\metre}$ rather than the $1\,\si{\milli\metre}$ thickness reported by Ramani et al.~\cite{ramaniSmartScanIntelligentScanning2022}.

\begin{figure}[htbp]
    \centering
    \includegraphics[width=0.55\textwidth]{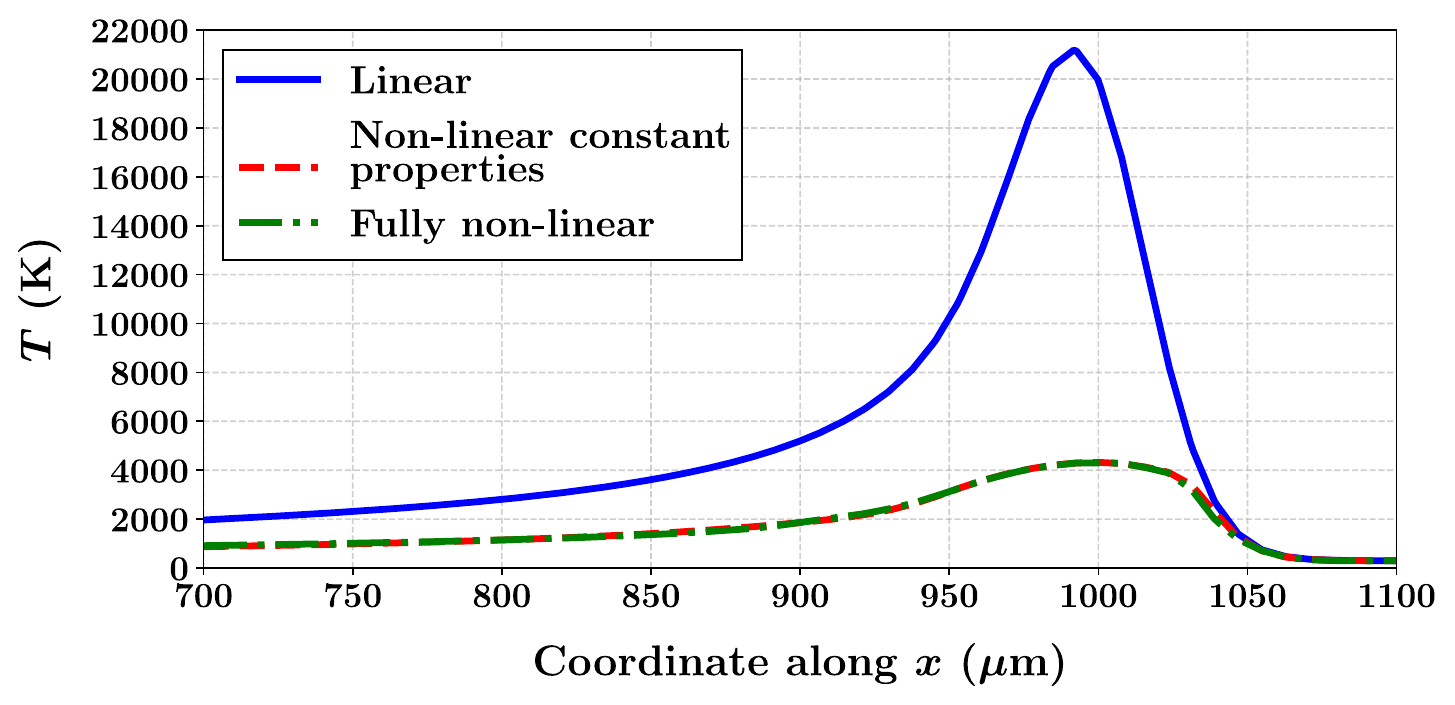}
    \caption{Centerline temperature on the top surface for a single laser pass, at three levels of physical fidelity: the linear physics of the case study~\cite{ramaniSmartScanIntelligentScanning2022}, the nonlinear case with constant material properties, and the fully nonlinear case with temperature-dependent properties. The three curves share the same refined grid.}
    \label{fig:effect_of_nonlinearities}
\end{figure}

Next, the role of nonlinearities on the thermal uniformity metric $R(t)$ is studied using the SG model.
Prior to reperforming the scanning strategies with nonlinearities, a single laser pass study similar to the one in section~\ref{sec:results_fe} is performed,  
with a refined grid in every case, and each time retaining the physics of   
(i) the benchmark linear case \cite{ramaniSmartScanIntelligentScanning2022}, (ii) the nonlinear case but with constant thermophysical properties, and (iii) the fully nonlinear case.
Figure~\ref{fig:effect_of_nonlinearities} shows the top surface line plot along the centerline of the melt pool for these three simulations. 
It illustrates that refining the grid from \SI{200}{\micro\meter} to \SI{50}{\micro\meter} while using the same \SI{77}{\micro\metre} beam diameter of \cite{ramaniSmartScanIntelligentScanning2022} significantly raises the peak temperature  from \SI{1667.3}{\kelvin} reported in \cite{ramaniSmartScanIntelligentScanning2022} to $21186$~K (Figure~\ref{fig:effect_of_nonlinearities}).
Activating the contributions of latent heat of fusion and evaporation then significantly reduce this peak temperature; 
evaporation removes a large share of the absorbed energy and reshapes how heat accumulates in the domain.
Meanwhile, the nonlinearity in thermophysical properties is found to play a negligible role for the specific 
laser conditions when both evaporation and latent heat of fusion are active.

\begin{figure}[htbp]
    \centering
    {\raggedright\hspace{1.9cm}\begin{subfigure}[b]{0.70\textwidth}
        \centering
        \includegraphics[width=\textwidth]{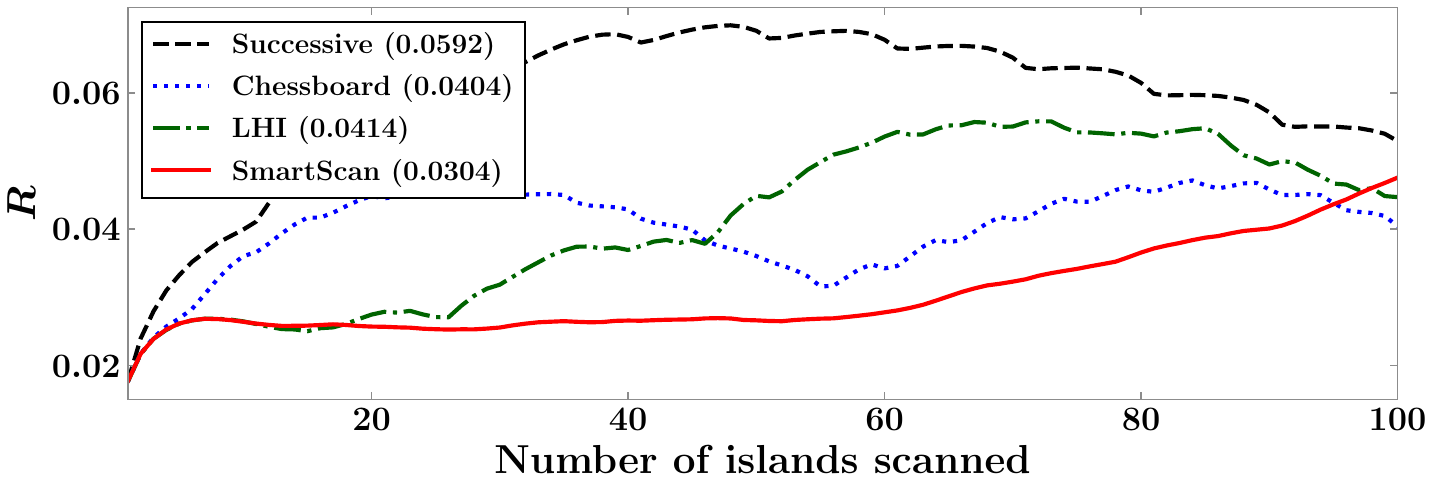}
        \caption{Thermal uniformity metric $R$}
        \label{fig:ramani_refined_a}
    \end{subfigure}\par}
    \vskip\baselineskip
    \begin{subfigure}[b]{0.70\textwidth}
        \centering
        \includegraphics[width=\textwidth]{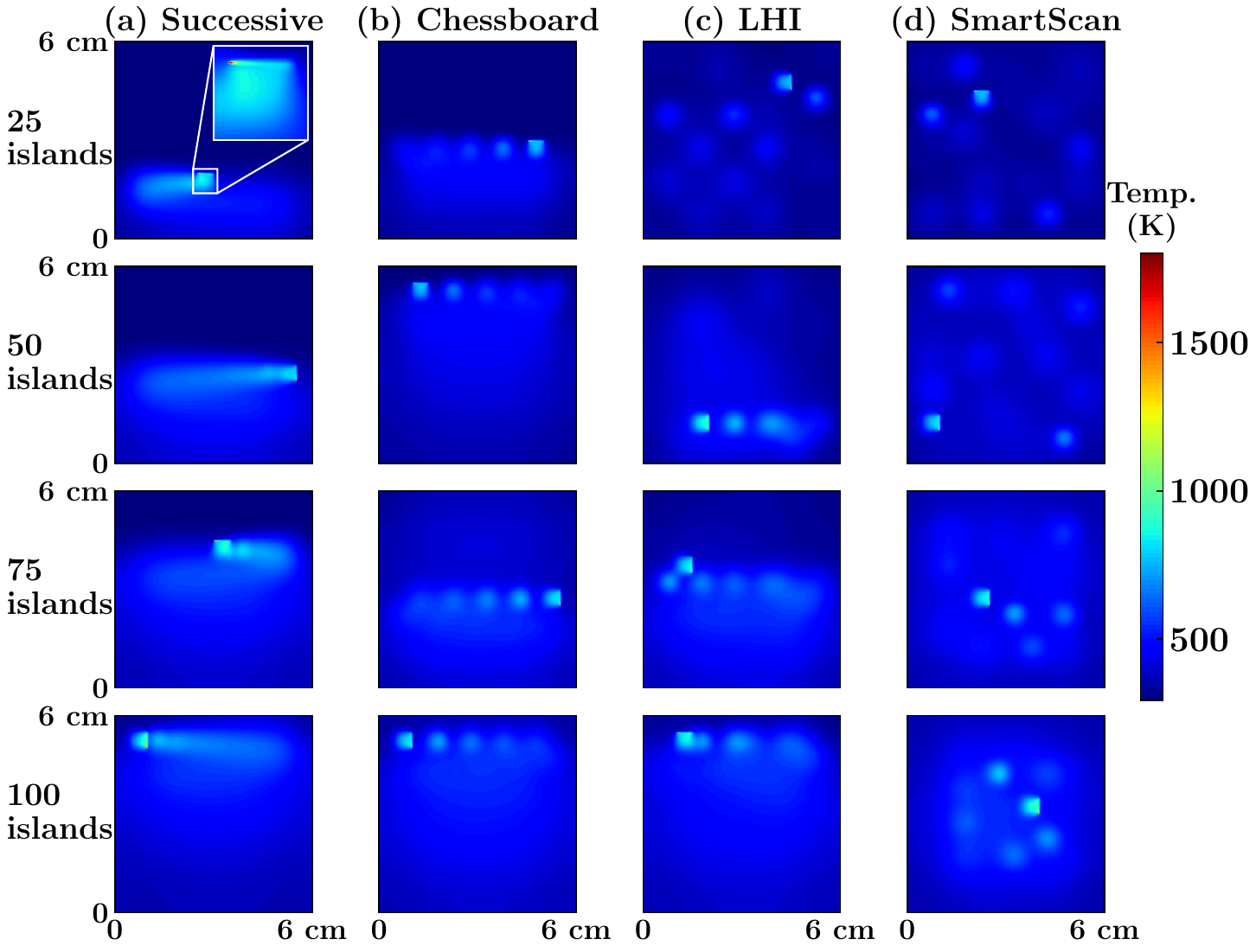}
        \caption{Temperature field}
        \label{fig:ramani_refined_b}
    \end{subfigure}
    \caption{Refined nonlinear SG simulation of the Ramani et al.~\cite{ramaniSmartScanIntelligentScanning2022} case study, run at $N_x = N_y = 1200$, $N_z = 8$ with evaporation and latent heat. (a) Uniformity metric $R$ versus number of islands scanned; the legend gives each strategy's mean $R$. (b) Temperature field after 25, 50, 75, and 100 islands for the four strategies, with an inset magnifying one island.}
    \label{fig:ramani_refined}
\end{figure}

The comparison above shows that grid spacing and latent heat and evaporation govern the peak temperature. 
The uniformity metric \eqref{uniformity_metric} directly depends on the temperature field, so the baseline model's omission of these nonlinearities may result in inaccurate ranking of the scan strategies.
Additionally, the grid spacing of \SI{200}{\micro\metre} used in \cite{ramaniSmartScanIntelligentScanning2022} is larger than the \SI{77}{\micro\metre} beam diameter, and it consequently diminishes the temperature peak, affecting in turn the strongly nonlinear evaporation.
The four scanning strategies are thus repeated with evaporation and latent heat enabled, keeping the thermophysical properties constant, and the grid is refined to $N_x = N_y = 1200$ and $N_z = 8$.
Figures~\ref{fig:ramani_refined_a} and~\ref{fig:ramani_refined_b} show the results of this simulation.

The addition of nonlinearities and grid refinement has a significant effect on the results. 
Relative to Fig.~\ref{fig:ramani_rep_a}, the thermal uniformity curves of Fig.~\ref{fig:ramani_refined_a} shifts down considerably but show a negligible change in the trend, and the temperature field diverges more sharply from the baseline: heat accumulation is significantly reduced across the domain due to evaporation, and the high temperature regions stay more confined to the islands that have been scanned most recently.
SmartScan continues to result in the most uniform temperature distribution; however, its thermal uniformity metric $R(t)$ is now significantly reduced.
Resolving the essential nonlinearities is therefore very important for this kind of study.
The results highlight the importance of higher-fidelity SG simulations in such studies to improve scan strategies.
Furthermore, the SG method solved this nonlinear run in \SI{15.3}{\hour} on the same hardware that solved the coarser linear baseline case of Fig.~\ref{fig:ramani_comparison} in \SI{36.7}{\minute}: approximately $25$ times longer; however, the significant gain in accuracy largely compensates for the computational time.

\section{Discussion}\label{sec:discussion}

The fundamental difference between the proposed SG method and the conventional FE approach lies in the choice of the orthonormal basis functions $\Phi_{mnp}$; in the SG method, the temperature is expanded over a complete trigonometric orthonormal basis that reduces the Galerkin projection to a set of modal ODEs in time, whereas the conventional Galerkin FE approach uses a polynomial basis that does not yield such a reduction.
Consequently, an implicit FE step incurs the additional computational cost of a global algebraic solve that the SG method avoids.
Instead, the SG method integrates the resulting modal ODEs using an exponential time-differencing scheme that reduces to a mode-wise multiplication in the spectral space.
This explains why the SG solver can be efficiently implemented on a GPU and runs nearly three orders of magnitude faster than the FE simulation with first-order Lagrange elements, with a negligible impact on accuracy.

For the specific cases of laser scanning studied in this work, the convergence of the SG approach is found to be the slowest in the direction normal to the laser scanning surface i.e., the vertical direction; $N_z$ larger than $N_x$ and $N_y$ was needed to converge.
This discrepancy arises from the manner in which the boundary condition is treated.
Each mode has a zero derivative on the top surface, so the trigonometric basis cannot represent the non-zero laser and evaporative flux due to laser scanning on the top surface directly; the flux enters only through the surface forcing, whose vertical coefficients decay as $p^{-2}$. The volumetric field then converges as $N_z^{-3/2}$, and the inhomogeneous surface temperature only as $N_z^{-1}$. 
Therefore, the top surface sets the requirement for the resolution in this direction. 
Lifting the flux into a smooth particular solution could restore the homogeneous condition and recover fast convergence.

The SG framework for temperature field resolution involving melting does not resolve melt pool dynamics. 
Handling melt-pool convection, Marangoni flow, and recoil or keyhole effects requires solving the Navier-Stokes equation with its own eigenbasis along with the heat transfer problem.
Another possible approach could be a two-scale coupling between a finely resolved melt-pool dynamics simulation on a subdomain and a coarser far-field heat transfer resolution on the larger surrounding domain with a shared Neumann interface in between; each problem would be independently integrated at its own resolution.
This multiscale coupling could be enabled by the fact that the current fixed-domain SG framework carries flux data on its faces. 

Evolving the domain, for example, via simulating material addition during additive manufacturing would require projecting the temperature solution onto the eigenbasis of the enlarged domain at each deposition step; the properties of the new layer then would enter the forcing term.
The same property correction could also admit arbitrary geometries; for example, the final arbitrary-shape of the part-to-be-built could sit inside a bounding cuboid with voxel activation as building proceeds, similar to element activation/deactivation that is now standard practice in FE-based simulations for additive manufacturing applications.

Finally, the explicit lagging within a time step imposes a convergence restriction on the forcing that becomes stronger as the nonlinearity strengthens. 
The Picard (fixed-point) iterations converge linearly and the convergence is more gradual where the nonlinearity is strong; this can be resolved using better iterative schemes such as Anderson acceleration \cite{andersonIterativeProceduresNonlinear1965, walkerAndersonAccelerationFixedPoint2011}, which reuses a few previous iterates to form a better estimate without a Jacobian. 
Such a scheme should reduce the iteration count and enlarge the convergence domain of the strongly temperature-dependent regime. 

\section{Conclusion}

This work presents a semi-analytical spectral Galerkin (SG) framework for the efficient solution of nonlinear transient heat transfer problems in finite domains. 
The central idea is to separate the governing problem into a linear isotropic reference operator with an analytical eigensystem and residual forcing terms containing the nonlinear physics. 
Galerkin projection on this basis diagonalizes the reference operator and generates a system of modal ODEs in time.
Each modal contribution is advanced analytically using exponential time differencing, while the nonlinear forcing is resolved iteratively in physical space.
The resulting formulation avoids the repeated global algebraic solves of an implicit FE approach while retaining temperature-dependent properties, anisotropic conductivity, latent heat, and nonlinear volumetric and surface fluxes.

Applied to study rapid laser--metal interactions with melting, the SG predictions agree with the Eagar--Tsai solution to within 0.1\% in the linear regime and with a highly resolved nonlinear FE simulation to within 0.67\% over the domain. 
On the same single-core hardware, the SG solver is 6.2 times faster than FE simulations despite using approximately one order of magnitude more degrees of freedom.
However, its structured-grid formulation provides a significant speed-up as it unlocks efficient GPU parallelization, reducing the nonlinear simulation time from 189 h for the single-core FE reference to 0.83 h, corresponding to a 227-fold speed-up.

The computational gain makes it possible to retain nonlinear physics in problems for which high-fidelity FE simulations become impractically expensive. 
A published part-scale scan-strategy problem \cite{ramaniSmartScanIntelligentScanning2022} is revisited and it shows that accounting for latent heat and evaporation more than 
halves the thermal-uniformity metric in that work, while preserving the ranking of the scan strategies.
Linear thermal models may therefore reproduce qualitative trends while substantially altering the quantitative energy balance 
on which process metrics are built. 
Further improvements through treatment of nonhomogeneous surface fluxes and accelerated nonlinear iterations should extend the efficiency of the method to still larger and more strongly nonlinear problems. 
More broadly, the framework bridges the gap between computational efficiency of analytical approaches and the fidelity of high-resolution FE-based numerical methods for nonlinear heat transfer in finite domains.

\section*{Acknowledgements}

\noindent This work was partially supported by the Agence de l'innovation de d\'{e}fense -- AID -- via Centre Interdisciplinaire d'Etudes pour la D\'{e}fense et la S\'{e}curit\'{e} – CIEDS (project 2025 - LaserSurf), and by the French Agence Nationale de la Recherche -- ANR -- under grant number ANR-24-CE08-3737.

\section*{CRediT author statement}

\noindent \textbf{Théo Andrieux}: Conceptualization, Methodology, Software, Validation, Formal Analysis, Investigation, Data Curation, Writing -- Original Draft, Writing -- Review \& Editing, Visualization

\noindent \textbf{Andreas Ntinos}: Methodology, Software

\noindent \textbf{Manas V. Upadhyay}: Conceptualization, Methodology, Validation, Resources, Writing -- Original Draft, Writing -- Review \& Editing, Supervision, Project administration, Funding acquisition

\bibliographystyle{elsarticle-num}
\bibliography{references}

\appendix

\section{Algorithm for the implicit FE solver}
\label{appendix:impl_fe}

For reproducibility, the main implementation workflow of the finite-element solver used for reference comparisons is summarized in Algorithm~\ref{alg:implicit_fe_solver}.

\begin{algorithm}[H]
    \caption{Implicit FE Solver (FEniCSx)}
    \label{alg:implicit_fe_solver}
    \begin{algorithmic}[1]\small
        \STATE \textbf{Preprocessing:}
        \STATE Read the mesh; define the FE discrete finite-dimensional subspace $V_h$
        \STATE Set parameters
        \STATE \textbf{Time Integration Loop:}
        \FOR{each time step $\Delta t$}
            \STATE Update the laser flux $q_{\mathrm{las}}^{n+1}$ on $\partial \Omega$
            \STATE Set Newton initial guess $T^{n+1,0} \gets T^n$

            \STATE \COMMENT{--- Nonlinear solve at step $n+1$ ---}
            \FOR{Newton iteration $m$}
                \STATE Assemble residual $R(T^{n+1,m})$ (transient, diffusion, laser, latent, heat loss terms)
                \STATE Assemble Jacobian $J(T^{n+1,m}) = \partial R / \partial T$
                \STATE Solve linear system $J\,\delta T = -R$
                \STATE Update $T^{n+1,m+1} \gets T^{n+1,m} + \delta T$
                \IF{nonlinear convergence achieved (e.g., $\|R\| < \text{tol}$)}
                    \STATE \textbf{break}
                \ENDIF
            \ENDFOR

            \STATE Write converged solution: $T^{n+1} \gets T^{n+1,m+1}$
        \ENDFOR
    \end{algorithmic}
\end{algorithm}

\begin{figure}[ht!]
    \centering
    \includegraphics[width=0.8\textwidth]{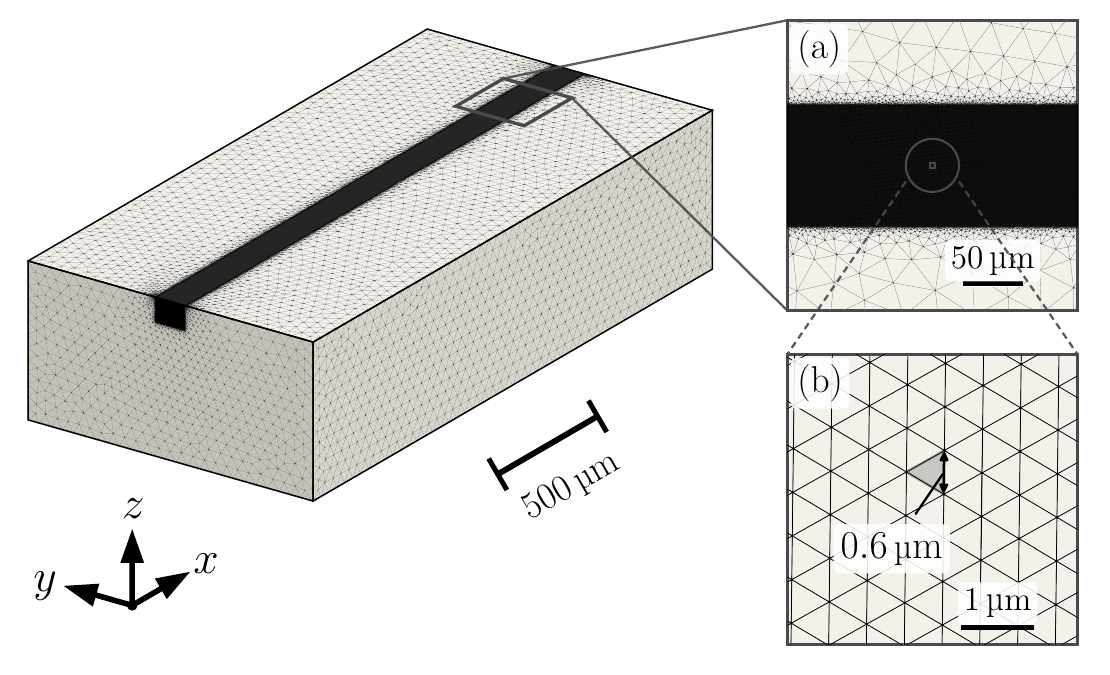}
    \caption{Mesh used for the FE simulations, linear tetrahedra with a coarse background of $40$~\si{\micro\metre} and insets showing (a) the refined region along the laser trajectory and (b) the finer $0.6$\si{\micro\metre} tetrahedra mesh. This size was selected by a convergence study.} 
    \label{fig:mesh_fe}
\end{figure}

\end{document}